\documentclass[useAMS,usenatbib]{mn2e}
\input psfig.tex

\usepackage{latexsym}
\usepackage{amssymb}
\usepackage{amsmath}
\usepackage{graphics}
\usepackage{graphicx}
\usepackage{natbib}
\usepackage{subfigure}

\usepackage{bm} 

\title[Theory of stellar convection: Removing the Mixing-Length Parameter ]{Theory of stellar convection:\\
 Removing the Mixing-Length Parameter}
\author[S. Pasetto, C. Chiosi, M. Cropper, and E. K. Grebel]{S. Pasetto $^{1}$\thanks{E-mail:
s.pasetto@ucl.ac.uk}, C. Chiosi $^{2}$, M. Cropper $^1$, and E. K. Grebel $^{3}$ \\
 $^{1}$University College London, Department of Space \& Climate Physics, Mullard Space Science Laboratory, Holmbury St. Mary, \\ \,\,\,Dorking Surrey, United Kingdom\\
 $^{2}$Department of Physics \& Astronomy, "Galileo Galilei", University of Padua, Padova, Italy\\
 $^{3}$Astronomisches Rechen-Institut, Zentrum f\"ur Astronomie der Universit\"at Heidelberg, M\"onchhofstr. 12-14, 69120, \\
 \,\,\,Heidelberg, Germany\\   }

\begin{document}

\date{Accepted 2014 September 12. Received in original form 2014 March 24}

\pagerange{\pageref{firstpage}--\pageref{lastpage}} \pubyear{2014}

\maketitle

\label{firstpage}

\begin{abstract}
   {Stellar convection is customarily described by Mixing-Length Theory, which makes use of the mixing-length scale to express the convective flux, velocity, and  temperature gradients of the convective elements and stellar medium. The mixing-length scale is taken to be proportional to the local pressure scale height, and the proportionality factor (the mixing-length parameter) must be determined by comparing the stellar models to some calibrator, usually the Sun. No strong arguments exist to suggest that the mixing-length parameter is the same in all stars and at all evolutionary phases. }
   {The aim of this study is to present a new theory of stellar convection that does not require the mixing length parameter. We present a self-consistent analytical formulation of stellar convection that determines the properties of stellar convection as a function of the physical behaviour of the convective elements themselves and of the surrounding medium.}
   {This new theory is formulated starting from a conventional solution of the Navier-Stokes/Euler equations, i.e. the Bernoulli equation for a perfect fluid, but expressed in  a non-inertial reference frame co-moving with the convective elements. In our formalism the motion of stellar convective cells inside convectively-unstable layers is fully determined by a new system of equations for convection in a non-local and time-dependent formalism. }
   {We obtain an analytical, non-local, time-dependent sub-sonic solution for the convective energy transport that does not depend on any free parameter. The theory is suitable for the outer convective zones of solar type stars and stars of all mass on the main sequence band. The predictions of the new theory are compared with those from the standard mixing-length paradigm for the most accurate calibrator, the Sun, with very satisfactory results. }
   {}
\end{abstract}



\section{Introduction}\label{Introduction}
In stellar interiors  convection  plays an important role: together with radiation and conduction, it transports energy throughout a star, and it chemically homogenizes the regions affected by convective instability. Therefore convection significantly affects the structures and evolutionary histories of stars. For example, the centre of main sequence stars slightly more massive than the Sun and above is dominated by convective transport of energy. In stars less massive than about 0.3 $M_\odot$ the whole structure becomes fully convective. The outer layers of stars of any mass are convective toward the surface. Very extended convective envelopes exist in red-giant-branch (RGB) and asymptotic-giant-branch (AGB) stars. Pre-main sequence stars are fully convective along the Hayashi-line. Finally  convection is present in the pre-supernova stages of type I and II supernovae, and even during the collapse phase of type II supernovae \citep[e.g.][]{2014AIPA....4d1010A, 2014ApJ...785...82S, 2011ApJ...741...33A, 2007ApJ...667..448M}.
In most cases, convection in the cores and inner shells does not pose serious difficulties to our understanding of the structure of the stars because the large thermal capacity of convective elements results in the degree of ``super-adiabaticity'' being so small that for any practical purpose the temperature gradient of the medium in the presence of convection can be set equal to the adiabatic value, unless evaluations of the velocities and distances traveled by convective elements are required, e.g. in presence of convective overshooting \citep[see for instance the early studies by][]{1975A&A....40..303M,1975A&A....43...61M,1981A&A...102...25B}. Describing convection in the outer layers of a star is by far more difficult and uncertain. Convective elements in this region have low thermal capacity, so that the super-adiabatic approximation can no longer  be applied, and the temperature gradient of the elements and surrounding medium must be determined separately to exactly know the amount of energy carried by convection and radiation \citep[e.g.][]{ 2013sse..book.....K, 2004cgps.book.....W}.

A suitable description of convection is therefore essential to determine stellar structure. The universally adopted solution is the Mixing-Length Theory (MLT) of convection, a simplified analytical formulation of the problem. Unfortunately, a more satisfactory analytical treatment of stellar convection is still missing and open to debate \citep[e.g.][]{2011A&A...528A..76C}. The MLT stands on the works of \citet[][]{1951ZA.....28..304B} and \citet[][]{1958ZA.....46..108B} which are based on earlier works on the concept of convective motion by \citet{Prandtl}. In this standard approach, the motion of convective elements is related to the mean-free-path $l_m$ that a generic element is \textit{supposed} to travel at any given depth inside the convectively unstable regions of a star \citep[e.g.][Chapter 7]{ 1994sse..book.....K}. The mean free path $l_m$ is assumed to be proportional to the natural distance scale $h_P$  given by the pressure stratification of the star. The proportionality factor is however  poorly known and constrained. The mixing-length (ML) parameter $\Lambda _m$, defined by $l_m \equiv {\Lambda _m}{h_P}$, must be empirically determined. Nevertheless, the knowledge of this parameter is of paramount importance in correctly determining the convective energy transport, and hence the radius and effective temperature of a star.
This critical situation explains the many versions of convection theory that can be found when investigated in different regions and evolutionary phases of a star such as the overshooting from core or envelopes zones \citep[e.g.][]{ 2008MNRAS.386.1979D, 2007A&A...475.1019C,1981A&A...102...25B}, the helium semi-convection in low and intermediate mass stars $m < 5{M_ \odot }$ \citep[e.g.][]{ 1993A&AS..100..647B, 1985ApJ...296..204C}, the time-dependent convection in the carbon deflagration process in Type I supernovae  \citep[e.g.][]{ 1976Ap&SS..39L..37N}, the studies on the efficiency of convective overshooting \citep[e.g.][]{ 2013arXiv1301.7687B}, and the effects of rotation \citep[e.g.][]{2008A&A...479L..37M} to mention just a few.

Examining the classical formulation of the MLT presented in any textbook,
see for instance \citet[][]{Hofmeister1964}, \citet[][]{1968pss..book.....C} and their modern versions
\citep[][respectively]{2013sse..book.....K,2004cgps.book.....W}, we note that the MLT reduces to the energy conservation principle supplemented by an estimate of the mean velocity of convective elements.  In a convective region the total energy flux ($\varphi$) is the sum of the convective flux ($\varphi_{\rm{cnv}}$) and the radiative flux ($\varphi_{\rm{rad}}$); the total flux  is set proportional to a fictitious radiative gradient $\nabla_{\rm{rad}}$ \footnote{Throughout the paper, we will introduce several   logarithmic temperature gradients  with respect to pressure $\frac{{d\log T}}{{d\log P}}$, shortly indicated as $\nabla$. Each of these gradients is also  identified by a subscript such as ${\nabla _e}$, ${\nabla _{\bm{\xi }}}$, ${\nabla _{{\rm{ad}}}}$, $\nabla_{\rm{rad}}$ depending of the circumstances. Finally,  the symbol $\nabla$ with no subscript is reserved for the ambient temperature gradient with respect to pressure across a star. }
(which is always known once the total flux coming from inside is assigned, typically case in stellar interiors); the true  radiative flux $\varphi_{\rm{rad}}$ is proportional to the real gradient of the medium $\nabla$; and the convective flux $\varphi_{\rm{cnv}}$ is proportional to the difference between the gradient of the convective elements and the gradient of the medium ($\nabla_e - \nabla$). By construction, the convective flux is  also proportional to the mass of an ideal convective element, i.e., the amount of matter crossing the unit area per unit time with the mean  velocity of convective elements. These elements may have any shape, mass,  velocity and lifetime, and may travel different distances before dissolving into the surrounding medium, releasing their energy excess and inducing mixing in the fluid. However all this ample variety of possibilities is simplified to an ideal element of averaged dimensions, lifetime, mean velocity and distance travelled before dissolving: the so-called mixing-length $l_m$ (and associated mixing-length parameter $\Lambda_m$). As far as the velocity is concerned, this is estimated from the work done by the buoyancy force over the distance $l_m$, a fraction of which is supposed to go into kinetic energy of the convective elements. Since in this problem the number of unknowns exceeds the number of equations (flux conservation and velocity), two more suitable relations are usually added. These are  firstly the ratio between the excess of energy in the bubble just before dissolving, to the energy radiated away (lost) during the lifetime, and secondly the excess rate of energy generation minus the excess rate of energy loss by radiation in the element relative to the surroundings. These are all functions of $\nabla$,  $\nabla_e$ and $\nabla_{\rm{ad}}$, see e.g. \citet[][]{1968pss..book.....C}. Now the number of unknowns, i.e. $\varphi_{\rm{rad}}$, $\varphi_{\rm{cnv}}$, $\nabla$, $\nabla_e$, is equal to the number of equations and the problem can be solved once $l_m$ or $\Lambda_m$ are assigned. \textit{In this way the complex fluid-dynamic situation is reduced to an estimate of the mean element velocity simply derived from the sole buoyancy force, neglecting other fluid-dynamic forces that can shape the motion of convective elements as function of time and surrounding medium}.

We present here a  new description of stellar convection that provides a simple and yet dynamically complete fully analytical integration of the hydrodynamic equations, matching the existing literature results based on the classical MLT, but without making use of any mixing-length  parameter ${\Lambda _m}$.

The plan of the paper is as follows. In Section \ref{probIntro2} we formulate the problem within the mathematical framework we intend to adopt. In Section \ref{Velpot} we define the concept of a scalar field of the velocity potential for expanding/contracting convective elements.  In Section \ref{EqcnvQ} and \ref{Accele} we formulate the equation governing the two degrees of freedom of our dynamical system: In Section \ref{EqcnvQ} formulates the equation of motion for a convective element as seen by a non-inertial frame of reference co-moving with it, and presents two lemmas that are functional  to our aim; in Section \ref{Accele} we solve the equation of motion of a convective element expressed in the co-moving frame of reference.
In Section \ref{Results} we present the predictions of our theory. First, we formulate the basic equations of stellar convection showing that the mixing-length parameter is no longer required.  Then we apply the new formalism to the case of the Sun. Finally in Section \ref{Conc} we present some concluding remarks highlighting the novelty and the power of the new theory.

\section{Formulation of the problem}\label{probIntro2}
Inside a  convective unstable layer, upward (downward) displacements of convective cells continually occur. The upwardly displaced elements are hotter and lighter than their surroundings, at the same pressure,  so that heat exchange results in energy release to the surrounding interstellar medium. For downward displacements the result is the opposite: convective cells sink when they have lower temperatures than their surroundings, and are heated on some length scale. A mathematical formalism for this process is presented in Section \ref{convflux}. Here we focus on the motion of a single convective element.

Our starting point is represented by the classical solution of the Navier-Stokes equations for an incompressible perfect fluid where no electromagnetic forces are taken into account \citep[e.g.][]{ 1961hhs..book.....C}. We approach the mechanics of the convection by approximating the stellar fluid as a perfect fluid, i.e. a fluid of density $\rho $ in which a suitable equation of state (EoS)  links density $\rho  = \rho \left( {P,T,\mu } \right)$ with pressure $P = P\left( {{\bm{x}};t} \right)$, temperature $T = T\left( {{\bm{x}};t} \right)$ and molecular weight $\mu  = \mu \left( {{\bm{x}};t} \right)$ at a given instant $t$ and position  ${\bm{x}}$  inside a star (see also footnote 3). Perfect fluids are intrinsically unstable and turbulent, therefore the higher the Reynolds numbers  characterizing the fluid the better the above approximation holds.  It is well known \citep[e.g.][]{ 1961hhs..book.....C} that in stellar interiors where turbulence prevails over viscosity in the Navier-Stokes equations, the inertial term $\left\langle {\rho {{\bm{v}}_0},{\nabla _{\bm{x}}}{{\bm{v}}_0}} \right\rangle $ dominates over the viscous one, $ - \eta {\underline \Delta  _{\bm{x}}}{{\bm{v}}_0}$. Here ${{\bm{v}}_0}$ is the stellar fluid velocity,  $\eta $ the viscosity coefficient,  $\left\langle {*,*} \right\rangle $ the standard inner-product between two generic vectors, and ${\nabla _{\bm{x}}}$ and ${\underline \Delta  _{\bm{x}}}$ the gradient and Laplacian operators, respectively, for an inertial reference system of coordinate ${S_0}\left( {O,{\bm{x}}} \right)$
centred in $O$ at the centre of the star with direction vector $\bm{x }$. Then, if in the equation of motion (EoM) for the stellar plasma we neglect the contribution of the magnetic field ${\bm{B}}$, i.e. the term $\frac{{\bm{j}}}{c} \times {\bm{B}}$ where ${\bm{j}} = \rho {{\bm{v}}_0}$ is the charge-current-density (and $* \times *$ is the cross-product between two generic vectors) the corresponding Euler's equation, $\frac{{\partial \rho {{\bm{v}}_0}}}{{\partial t}} + \left\langle {\nabla ,{\bm{P}} + \rho {{\bm{v}}_0}{{\bm{v}}_0}} \right\rangle  - \rho \sum\limits_i^{} {{n_i}{{\bm{F}}_i}}  = 0$, together with the continuity equation, $\frac{{\partial \rho }}{{\partial t}} + \left\langle {\nabla ,\rho {{\bm{v}}_0}} \right\rangle  = 0$ and accounting for the relation $\left\langle {\nabla ,\rho {{\bm{v}}_0}{{\bm{v}}_0}} \right\rangle  = {{\bm{v}}_0}\rho \nabla  \cdot {{\bm{v}}_0} + \rho {{\bm{v}}_0} \cdot \nabla {{\bm{v}}_0}$, reads:

\begin{equation}\label{Eq001}
\rho \frac{{\partial {{\bm{v}}_0}}}{{\partial t}} + \left\langle {{\nabla _{\bm{x}}},{\bm{P}}} \right\rangle  + \left\langle {\rho {{\bm{v}}_0},{\nabla _{\bm{x}}}{{\bm{v}}_0}} \right\rangle  - \sum\limits_i^{} {{n_i}{{\bm{F}}_i}}  = 0.
\end{equation}

\noindent In  Eq.(\ref{Eq001}),  ${\bm{P}}$ is the pressure tensor, ${\bm{F}}$ the force acting on every particle of the fluid, ${n_i}$ the number density of every type of fluid particle (with the above assumption that no electric field ${\bm{E}}$ enters the plasma EoM).
This is a partial differential equation (PDE) where the quantities involved, say $Q$, are functions of time $t$ and position ${\bm{x}}$, $Q=Q(\bm{x};t)$ in the given inertial reference frame ${S_0}\left( {O,{\bm{x}}} \right)$. Hereafter, we omit writing this dependence explicitly to simplify the notation (unless specified otherwise for the sake of better understanding).
Stellar interiors on macroscopic scale are well represented by a perfect fluid in local  thermodynamical equilibrium (LTE), i.e. each elemental component, $n_i$ of the fluid is isotropic, homogeneous, in mechanical equilibrium and obeying the conditions of detailed balance with any other component $n_j$.
Therefore, we can then simplify the pressure tensor to a scalar $\left\langle {{\nabla _{\bm{x}}},{\bm{P}}} \right\rangle  = {\nabla _{\bm{x}}}P$, and because the force acting on the fluid particle is non-diffusive, i.e. in our case the gravity ${\bm{F_i}} = {m_i}{\bm{g}}$ on the particles of the $i$-th species, we assume that $\sum\limits_i^{} {{n_i}{\bm{F_i}}}  = \sum\limits_i^{} {{m_i}{n_i}{\bm{g}}}  = \left( {\sum\limits_i^{} {{m_i}{n_i}} } \right){\bm{g}} = \rho {\bm{g}}$. All this further simplifies Eq. (\ref{Eq001}) to:

\begin{equation}\label{Eq002}
	\rho \frac{{\partial {{\bm{v}}_0}}}{{\partial t}} + {\nabla _{\bm{x}}}P + \left\langle {\rho {{\bm{v}}_0},{\nabla _{\bm{x}}}{{\bm{v}}_0}} \right\rangle  - \rho {\bm{g}} = 0.
\end{equation}	

\noindent We proceed further with an additional simplification by assuming that the stellar fluid is incompressible and irrotational on  large distance scales. The concept of a large distance scale for incompressibility and irrotationality is defined  here from a heuristic point of view: This length should be large enough to contain a significant number of convective elements so that a statistical formulation is possible when describing the mean convective flux of energy (see below), but small enough so that the distance travelled by the convective element is short compared to the typical distance over which significant gradients in temperature, density, pressure etc. can develop (i.e. those gradients are locally small).
These assumptions stand at the basis of every stellar model integration in the literature, and are fully compatible with making use of the simple concept of a potential flow \citep[e.g.][Chapter 1]{1959flme.book.....L}:  ${\nabla _{\bm{x}}} \times {{\bm{v}}_0} = 0 \Leftrightarrow \exists {\Phi _{{{\bm{v}}_0}}}\, | \, \, \,{{\bm{v}}_0} = {\nabla _{\bm{x}}}{\Phi _{{{\bm{v}}_0}}}$ with $\Phi _{{{\bm{v}}_0}}$ the velocity potential. In particular, with the help of the vector relation $\left\langle {{{\bm{v}}_0},{\nabla _{\bm{x}}}{{\bm{v}}_0}} \right\rangle  = \frac{1}{2}{\nabla _{\bm{x}}}\left\langle {{{\bm{v}}_0},{{\bm{v}}_0}} \right\rangle  - {{\bm{v}}_0} \times \left( {{\nabla _{\bm{x}}} \times {{\bm{v}}_0}} \right)$ and remembering that the curl of a gradient is null, ${\nabla _{\bm{x}}} \times {{\bm{v}}_0} = {\nabla _{\bm{x}}} \times {\nabla _{\bm{x}}}{\Phi _{{{\bm{v}}_0}}} = 0$, we are able to write Eq. (\ref{Eq002}) as $
{\nabla _{\bm{x}}}\left( {\frac{{\partial {\Phi _{{{\bm{v}}_0}}}}}{{\partial t}} + \frac{P}{\rho } + \frac{{{{\left\| {{{\bm{v}}_0}} \right\|}^2}}}{2} + {\Phi _{\bm{g}}}} \right) = 0$ where the relation between gravitational force and gravitational potential ${\bm{g}} =  - {\nabla _{\bm{x}}}{\Phi _{\bm{g}}}$ has been adopted. The symbol $\left\| {*} \right\|$ indicates the standard Euclidian norm of a generic vector. Finally, the integration of the previous equation leads to the Bernoulli's equation in an inertial reference system ${S_0}$ centred at the centre of the star. With the formalism developed here:

\begin{equation}\label{Eq003}
	\frac{{\partial {\Phi _{{{\bm{v}}_0}}}}}{{\partial t}} + \frac{P}{\rho } + \frac{{{{\left\| {{{\bm{v}}_0}} \right\|}^2}}}{2} + {\Phi _{\bm{g}}} = f\left( t \right).
\end{equation}	

\noindent This is one of the basic equations describing the stellar plasma in which convection is at work. A more complete treatment would include diffusion  and turbulence. However, as the main goal here is to derive the mechanics of convection from simple principles, the present approach is adequate for our aims. Diffusion and turbulence can eventually be included using the same formalism in a future study.
In the context of thermal convection, it is worth recalling that the Boussinesq \citep{1960ApJ...131..442S} and anelastic \citep{1969JAtS...26..448G} approximations would be valuable alternatives  worth being investigated. Nevertheless, for the aims of this study the potential flow approximation turns out to be fully satisfactory at an extremely high degree of precision as our numerical investigation in Section \ref{NumEq} will confirm.

After these preliminary remarks, we are now in the position to state the queries that we intend to address as follows: the main target of  stellar convection is to find a solution of Eq.(\ref{Eq003})  linking  the physical quantities characterizing the stellar interiors  such as pressure, density, temperature, velocities etc. and the mechanics governing the motion  of the convective elements as functions of  the fundamental temperature gradients with respect to pressure, i.e. the radiative gradient ${\nabla _{{\rm{rad}}}}$, the adiabatic gradient ${\nabla _{{\rm{ad}}}}$, the local gradient of the star $\nabla  \equiv \left| {\frac{{d\ln T}}{{d\ln P}}} \right|$, the convective element gradient ${\nabla _e}$ and the molecular weight gradient ${\nabla _\mu } \equiv \left| {\frac{{d\ln \mu }}{{d\ln P}}} \right|$.

The task is difficult because of the large number of variables involved to describe the physics of the convective element and of the stellar interiors, both of which poorly known. Mathematically, the problem  translates into a system of Algebraic-Differential Equations (ADEs). In the MLT, the solution of this ADE is simplified to an algebraic system of equations by introducing a statistical description of the motion, size, lifetime etc. of  the convective elements. In this way, the complicated pattern of possible convective elements is reduced to a mean element whose dimensions and path are simply supposed to be $l_m=\Lambda_m  h_P$, where $\Lambda_m$ is a parameter to be fixed by comparing real stars (the Sun) to stellar models. Once $\Lambda_m$ is calibrated is this way, it is assumed to be the same for all stars of any mass, chemical composition, and evolutionary stage. This is indeed a strong assumption.

In what follows we propose and formulate an alternative approach to this theory and apply it to recover well-established results of the theory of stellar structure and observational properties of our best calibrator, the Sun, but without making use of any adjustable parameter.

The approach is based on the addition of an equation for the motion of the convective
elements to the classical system of algebraic equations for the convective energy transport. The whole system of algebraic/differential equations is solved by considering together the evolution of the generalized coordinates (i.e., independents or Lagrangian coordinates associated with the degree of freedoms) of the radius and position of a convective element.
This result is achieved by means of a series of theorems, corollaries and lemmas that
analyse the different physical and mathematical aspects of the problem.

\begin{figure}
\includegraphics[width=\columnwidth]{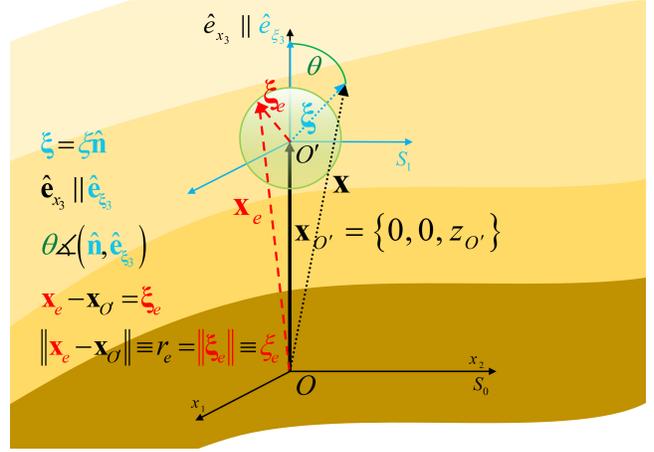}
\caption{Schematic representation of the quantities involved in  the two reference systems introduced in the text. The star is supposed to be stratified in hydrostatic equilibrium layers (each of the monochromatic background colours) and the convective elements are living inside these layers. In green is an
example of an expanding convective element. The reference frame  $S_0$ (in black) is fixed at the centre of the star (supposed to be at rest or in translational
motion), and the frame $S_1$ (in blue) is the non-inertial system of reference and it is located with the convective element along the z-axis of the $S_0$ system of reference, ${{\bm{x}}_{O'}} = \left\{ {0,0,{z_{O'}}} \right\}$.  The geometry employed is
completely general and derived from a simplified version of Fig. 1 of \citet{2009A&A...499..385P}. The position vectors of the two system of reference are in black for the system $S_0$ and blue for $S_1$ respectively; they take red colour only when they refer to surface of the sphere instantaneously matching the location and size of the convective element considered. In green it is the angle between the direction of motion and the $S_1$ position vector.
\label{Fig1} }
\end{figure}

Before starting our analysis, in order to avoid a possible misunderstanding of the real meaning of some of our analytical results, it might be wise to call attention to a formal aspect of the mathematical notation we have adopted. For some quantities $Q$ function of time or space or both, $Q({\mathbf{x}};t)$, we look at their asymptotic behaviour by formally taking the limits
\begin{equation}\label{Eq000}
	{Q^\infty } \equiv \mathop {\lim }\limits_{\scriptstyle {\bm{x}} \to {{\bm{x}}^\infty } \hfill \atop
		\scriptstyle t \to \infty  \hfill}  Q\left( {{\bm{x}};t} \right) = Q\left( {{{\bm{x}}^\infty };\infty } \right).
\end{equation}
\noindent This  does not mean that we are taking temporal intervals infinitely long, rather that we are considering time long enough so that the asymptotic trend of the quantity $Q$ is reached but still short enough so that the physical properties of the whole system have not changed significantly, such as that the star still exists. In analogy, with the notation ${{{\bm{x}}^\infty }}$ we refer to a location far away from the system considered (e.g. the convective element in consideration) but at a given location ${\bm{x}} = {{\bm{x}}^\infty }$ that is still \textit{inside} the star, i.e. where $\rho  = \rho \left( {{{\bm{x}}^\infty }} \right) \ne 0$, $T = T\left( {{{\bm{x}}^\infty }} \right) \ne 0$ etc.

\section{Velocity-potential scalar-field for expanding/contracting convective elements}\label{Velpot}
The formalism we will develop refers to stars in hydrostatic equilibrium under the effect of their own gravity\footnote{It has long been known that the very external regions of some types of stars, e.g. pulsating stars, may deviate from rigorous hydrostatic equilibrium. Furthermore, some stars may experience evolutionary phases far from hydrostatic equilibrium, e.g. the collapsing core of massive stars or the accretion phase of proto-stars. In both cases convection may set in. Although these situations would represent an interesting field of investigation, these objects are not considered in the present study.} which applies to the vast majority. We limit ourselves to consider the onset of convection either in central cores, intermediate shells and external envelopes in the same conditions usually described by the classical MLT.

In addition to the natural inertial reference frame $S_0$ whose origin is fixed at the centre of the star (at rest by definition), we now introduce a  non-inertial reference frame co-moving with a generic convective element. The new system is named   ${S_1}$: $\left( {O',{\bm{\xi }}} \right)$, origin and position vector respectively, to distinguish it from the inertial reference frame ${S_0}$. Even though at first glance, this approach may look awkward, because  the most intuitive way of thinking about the motion of a body (in our case a convective element) is the translational motion with respect to the static observer, we show that this way of thinking yields the desired mathematical expression for the EoM of the convective element and eventually allows us to eliminate the mixing-length parameter.

Assuming spherical symmetry, in ${S_0}$ we define the equation of a generic convective element of  radius  ${r_e}$ by means of  the time-dependent relation $\left\| {\bm{r}} \right\| - {r_e}\left( t \right) = 0$ because the element is expected to expand/contract and rise/sink during  its lifetime evolution, and where we indicated with $\left\| {\bm{r}} \right\| = \left\| {{\bm{x}} - {{\bm{x}}_{O'}}} \right\|$  the radius vector of the element centred on ${{\bm{x}}_{O'}}$ at the instant $t = \hat t$. In $S_1$ we identify a point in the surface of the convective element by ${{\bm{\xi }}_e}$, then the radius of the convective element is ${\xi _e} \equiv \left\| {{{\bm{\xi }}_e}} \right\|$. Note that ${\xi _e} = {\xi _e}\left( t \right)$.
This is shown in Fig.\ref{Fig1} where the relation between the position vector in $S_0$ and $S_1$ is shown.

We define two velocity potentials, as introduced in Section 1, denoted by $\Phi _{{{\bm{v}}_0}}^I$ and $\Phi _{{{\bm{v}}_0}}^{II}$. $\Phi _{{{\bm{v}}_0}}^I$ is defined to satisfy the Laplace equation ${\Delta _{\bm{x}}}\Phi _{{{\bm{v}}_0}}^I = 0$ in ${S_0}$ in the case when the element is at rest, and this has particular solution $\Phi _{{{\bm{v}}_0}}^I =  - \frac{{{{\dot r}_e}r_e^2}}{{\left\| {\bm{r}} \right\|}}$ \citep[e.g.][]{ 1966hydr.book.....L}. The potential velocity  ${\bm{v}}_0^I = {v_r}{{\bm{\hat e}}_r} = \frac{{\partial \Phi _{{{\bm{v}}_0}}^I}}{{\partial r}}{{\bm{\hat e}}_r} = \frac{\partial }{{\partial r}}\left( { - \frac{{{{\dot r}_e}r_e^2}}{{\left\| {\bm{r}} \right\|}}} \right){{\bm{\hat e}}_r} = \frac{{{{\dot r}_e}r_e^2}}{{{{\left\| {\bm{r}} \right\|}^2}}}{{\bm{\hat e}}_r}$. Here the dot indicates the time derivative and ${{\bm{\hat e}}_r}$ is the normal radial vector of a polar coordinate system  centred on ${{\bm{x}}_{O}}$. The boundary conditions for Dirichlet's problem are: (i) far away from the convective element  $\mathop {\lim }\limits_{\left\| {\bm{r}} \right\| \to \infty } {\bm{v}}_0^I = 0$, and (ii) the kinematic-boundary-conditions at the surface of the sphere the surrounding plasma cannot flow through it, i.e. the velocity has to be purely tangential so that velocity component locally perpendicular to the surface must be zero $\left\langle {{\bm{v}}_0^I,{{{\bm{\hat e}}}_r}} \right\rangle_{r_e}  = 0$.

When the convective element is moving with velocity ${\bm{v}}$ in ${S_0}$,  the kinematic boundary condition is replaced by the relative velocity between the fluid and the element in motion throughout the stellar plasma $\left\langle {{{\bm{v}}_0} - {\bm{v}},{{{\bm{\hat e}}}_r}} \right\rangle  = 0$ at $\left\| {\bm{r}} \right\| = {r_e}$ to get the more general result for the potential flow $\Phi _{{{\bm{v}}_0}}^{II}$, where $\Phi _{{{\bm{v}}_0}}^{II} =  - \frac{1}{2}\frac{{r_e^3}}{{{{\left\| {\bm{r}} \right\|}^2}}}\left\langle {{\bm{v}},{{{\bm{\hat e}}}_r}} \right\rangle $ (to distinguish  it from $\Phi _{{{\bm{v}}_0}}^{I}$) with velocity given by ${\bm{v}}_0^{II} = {\nabla _{\bm{x}}}\Phi _{{{\bm{v}}_0}}^{II} = \frac{{r_e^3}}{{2{{\left\| {\bm{r}} \right\|}^3}}}\left( {3\left\langle {{\bm{v}},{{{\bm{\hat e}}}_r}} \right\rangle {{{\bm{\hat e}}}_r} - {\bm{v}}} \right)$ \citep[e.g.][]{ 1966hydr.book.....L}.

We move now our point of view to the non-inertial reference system ${S_1}$ co-moving with the convective element and with its axes always aligned with ${S_0}$: i.e. ${\bm{x}} = {{\bm{x}}_{O'}} + {\bm{\xi }}$ to get $\Phi_{{{\bm{v}}_0}}^{\prime II} =  - \left\langle {{\bm{v}},{\bm{\xi }}} \right\rangle \left( {1 + \frac{1}{2}\frac{{\xi _e^3}}{{{{\left\| {\bm{\xi }} \right\|}^3}}}} \right)$ for the potential flow past the convective element. This  is obtained by simply superimposing a translational potential vector, say  $\Phi _{{{\bm{v}}_0}}^t =  - \left\langle {{\bm{v}},{\bm{\xi }}} \right\rangle $ to the classical potential vector of the flow past a sphere in ${S_0}$.
$\Phi^\prime$ indicates  the potential flow passing the sphere at rest in ${S_1}$ (centred in $O'$ located at ${{\bm{x}}_{O'}} = {{\bm{x}}_{O'}}\left( t \right)$ in ${S_0}$) and with a radius vector ${\bm{\xi }}\left( t \right) = {\bm{x}} - {{\bm{x}}_{O'}}\left(t \right)$. Now the fluid potential velocity is ${\bm{v'}}_0^{II} = {\nabla _{\bm{x}}}\Phi_{{{\bm{v}}_0}}^{\prime II} = \frac{{\xi _e^3}}{{2{{\left\| {\bm{\xi }} \right\|}^3}}}\left( {3\left\langle {{\bm{v}},{\bm{\hat n}}} \right\rangle {\bm{\hat n}} - {\bm{v}}} \right) - {\bm{v}}$ with the desired boundary condition $\mathop {\lim }\limits_{\left\| {\bm{\xi }} \right\| \to \infty } {\bm{v'}}_0^{II} = {{\bm{v}}^\infty } =- {\bm{v}}$ and ${\bm{\hat n}}$,  the direction of the vector ${\bm{\xi}}$,  given by ${\bm{\hat n}} = \frac{{\bm{\xi }}}{{\left\| {\bm{\xi }} \right\|}}$.
Moreover, thanks to the linear character of the Laplace equation ${\underline\Delta _{\bm{x}}}{\Phi ^{{\rm{tot}}}} = {\underline\Delta _{\bm{x}}}\sum\limits_i^{} {{\Phi ^i}}  = \sum\limits_i^{} {{\underline\Delta _{\bm{x}}}{\Phi ^i}}  = 0$ for any assigned scalar potential $\Phi $, we  lump together the potential flows $\Phi_{{{\bm{v}}_{\rm{0}}}}^{\prime {\rm{tot}}}\equiv\Phi_{{{\bm{v}}_{\rm{0}}}}^{\prime {I}}+\Phi_{{{\bm{v}}_{\rm{0}}}}^{\prime {II}}$ for moving and contracting/expanding elements in order to obtain in $S_1$ the total potential flow outside the element surface as:
\begin{equation}\label{Eq004}
\Phi^\prime =  - \left\langle {{\bm{v}},{\bm{\xi }}} \right\rangle \left( {1 + \frac{1}{2}\frac{{\xi _e^3}}{{{{\left\| {\bm{\xi }} \right\|}^3}}}} \right) - \frac{{{{\dot \xi }_e}\xi _e^2}}{{\left\| {\bm{\xi }} \right\|}},	
\end{equation}
where  $\Phi^\prime \equiv \Phi_{{{\bm{v}}_{\rm{0}}}}^{\prime {\rm{tot}}}$ for the sake of simplicity. The  corresponding velocity in ${S_1}$ is obtained again as before by computing the gradient ${{\bm{v'}}_0} \equiv {\bm{v'}}_0^{{\rm{tot}}} = {\nabla _{\bm{\xi }}}\Phi^\prime$:
\begin{eqnarray}
\label{Eq005}
{{{\bm{v^\prime}}}_0} &=& {\left. {\frac{{\xi _e^3}}{{2{{\left\| {\bm{\xi }} \right\|}^3}}}\left( {3\left\langle {{\bm{v}},{\bm{\hat n}}} \right\rangle {\bm{\hat n}} - {\bm{v}}} \right) - {\bm{v}} + \frac{{{{\dot \xi }_e}\xi _e^2}}{{{{\left\| {\bm{\xi }} \right\|}^2}}}{\bm{\hat n}}} \right|_{\left\| {\bm{\xi }} \right\| = {\xi _e}}} \nonumber \\
                      &=& {\left. {\frac{3}{2}\left( {\left\langle {{\bm{v}},{\bm{\hat n}}} \right\rangle {\bm{\hat n}} - {\bm{v}}} \right) + {{\dot \xi }_e}{\bm{\hat n}}} \right|_{\left\| {\bm{\xi }} \right\| = {\xi _e}}},
\end{eqnarray}
where in order to get the final expression we have  evaluated the equation at the surface of the convective element $\left\| {\bm{\xi }} \right\| = {\xi _e}$ and simplified it. It is simple to  check that this equation yields correct results at the element surface, $\left\| {\bm{\xi }} \right\| = {\xi _e}$ once written in spherical coordinates (with $\theta $ the angle between ${{\bm{\hat e}}_z}$ and  ${{\bm{\hat \xi}}}$). For the radial component centred on $O'$ we obtain $v_{0,r}^{\prime} =  - v\left( {1 - \frac{{\xi _e^3}}{{{{\left\| {\bm{\xi }} \right\|}^3}}}} \right)\cos \theta  + \frac{{{{\dot \xi }_e}\xi _e^2}}{{{{\left\| {\bm{\xi }} \right\|}^2}}}$ by computing the Laplacian. It follows from this that  $v_{0,r}^{\prime} = {\dot \xi _e}$ at $\left\| {\bm{\xi }} \right\| = {\xi _e}$ because in ${S_1}$ the convective element does not  move with respect to the fluid  and the only residual velocity  in the fluid the result of the expansion/contraction.  In contrast,  for the component $v_{0,\theta }^{\prime} = v\left( {1 + \frac{{\xi _e^3}}{{2{{\left\| {\bm{\xi }} \right\|}^3}}}} \right)\sin \theta $,   at $\left\| {\bm{\xi }} \right\| = {\xi _e}$
we get  $v_{0,\theta }^{\prime} = \frac{3}{2}v\sin \theta $  which is maximum at $\theta  = \left\{ {\frac{\pi }{2},\frac{{3\pi }}{2}} \right\}$. Moreover, for $\left\| {\bm{\xi }} \right\| \to \infty $ we get ${v^\prime_{0,r}} =  - v\cos \theta $ and ${v'_{0,\theta }} = v\sin \theta $ so that at the stagnation point, $\theta  = 0$, we obtain ${v'_{0,r}} =  - v$ and ${v'_{0,\theta }} = 0$ as required by construction of the boundary conditions in the integration of Eq.(\ref{Eq004}) (in ${S_1}$ the convective cell  is at rest  and the fluid flows along a  direction opposite to the actual motion of the element in   ${S_0}$).
Finally,  we compute  the time derivative of the potential of Eq.(\ref{Eq004}) for use in the sections below and we recall that the surface of the convective element $\left\| {\bm{\xi }} \right\| = {\xi _e}$. The derivative is tedious but straightforward, and after a little algebra, we get
\begin{equation}\label{Eq006}
	{\left. {\frac{{\partial \Phi '}}{{\partial t}}} \right|_{\left\| {\bm{\xi }} \right\| = {\xi _e}}} =  - \frac{3}{2}{\xi _e}\left\langle {{\bm{A}},{\bm{\hat n}}} \right\rangle  - \frac{3}{2}{\dot \xi _e}\left\langle {{\bm{v}},{\bm{\hat n}}} \right\rangle  - {\ddot \xi _e}{\xi _e} - 2\dot \xi _e^2,	
\end{equation}
where the relative acceleration of the two reference frames is indicated with ${\bm{A}}$. This quantity will be examined in detail in  Section \ref{Accele}. The above analysis has given us two basic relationships, i.e.  Eq.(\ref{Eq005}) and (\ref{Eq006}) that determine the EoMs of the convective elements to be presented and discussed below.

\section{Equation governing the expansion/contraction of a convective element in ${S_1}$}\label{EqcnvQ}
The goal of this Section is to prove a relation  connecting  the evolution of the expansion rate of the convective element to the upward/downward motion inside the star.
This important relation is obtained as a corollary of a more general theorem once two independent lemmas are considered.

\subsection{Pressure-radius relation for expanding-contracting spheres in a non-inertial reference frame}
We want to prove the existence of a relation connecting the two Lagrangian coordinates that describe our system: $\left\| {{{\bm{x}}_{O'}}} \right\|$, the location of the convective element and $\left\|{{\bm{\xi }}_e}\right\|$, the size of the convective element. Being them independent variables, the relation for which we are searching has to involve the physical quantities describing  the environment in a given reference frame, which we choose to be ${S_1}$.

Instead of the classical approach reviewed in Section \ref{Introduction} proceeding from the Euler equation to Bernoulli's equation given by Eq.(\ref{Eq003}), we start  from deriving  the pressure acting on a convective  element in the non-inertial reference system ${S_1}$ (defined in Section \ref{Velpot}). This  problem has been recently discussed by  \citet{2012A&A...542A..17P} based on a previous approach presented by \citet{2009A&A...499..385P} (see their Section 3.1) extended to include a Navier-Stokes fluid-dynamics equation treatment. However, the (simpler) version of  \citet{2012A&A...542A..17P} was  developed for plasmas of much  higher temperatures  than the typical ones in  the stellar interiors, i.e. for the hot coronal plasma of the Milky Way, and did not consider the temporal evolution of the inner border of the fluid\footnote{In $S_1$ a convective element can be identified either by  an ``external'' surface delimiting its volume   or  the inner border of the external fluid  containing the convective element itself. In this case we can speak also of an external border for the fluid,  typically at +$\infty $ or far away from the convective element \citep[e.g.][]{2000ifd..book.....B}.}. The formalism developed there can however be adapted to the case of convective elements, with the new velocity-potential $\Phi'$  and associated velocity already given in Eq.(\ref{Eq004}) and Eq.(\ref{Eq005}) respectively, as follows.

\textbf{Theorem: pressure-radius relation for an expanding/contracting sphere in an external environment.}
We prove that the pressure and radius temporal evolution of an expanding/contracting convective element retaining its spherical shape is related by the following equation
\begin{align}\label{Eq010}
&\frac{{{v^2}}}{2}\left( {\frac{9}{4}{{\sin }^2}\theta  - 1} \right) - v{{\dot \xi }_e}\frac{3}{2}\cos \theta + \left( {\frac{P}{\rho } + {\Phi _{\bm{g}}}} \right) = \nonumber\\
&+A{\xi _e}\left( {\frac{3}{2}\cos \theta  - \cos \phi } \right) + {{\ddot \xi }_e}{\xi _e} + \frac{3}{2}\dot \xi _e^2,
\end{align}
holding in $S_1$, where $A = \left\| {\bm{A}} \right\|$ is the norm of the acceleration,  $\phi $ the angle between the direction of motion of the fluid as seen from ${S_1}$ and the acceleration direction,  and  $\theta $ has been already introduced before as $\theta \angle (\bm{v}, \bm{\xi})$.

\textbf{Proof}: We start using  Eq.(7) of \citet{2012A&A...542A..17P} written with the notation here set out. This is based on the same  hypotheses (without the Young-Laplace treatment of the surface tension) and is
\begin{equation}\label{Eq007}
\frac{{\partial \Phi '}}{{\partial t}} + \frac{P}{\rho } + \frac{{{{\left\| {{{{\bm{v'}}}_0}} \right\|}^2}}}{2} = f\left( t \right) - {\Phi _{\bm{g}}} - \left\langle {{\bm{A}},{\bm{\xi }}} \right\rangle .	
\end{equation}
With the boundary condition  for the hydrostatic equilibrium $\frac{P}{\rho } =  - {\Phi _{\bm{g}}}$ of the star, in the limit of $\left\| {\bm{\xi }} \right\| \to \infty $, it is easy to prove that we can fix the arbitrary function to be $f\left( t \right) = \mathop {\lim }\limits_{\left\| {\bm{\xi }} \right\| \to \infty } \frac{{{{\left\| {{{{\bm{v'}}}_0}} \right\|}^2}}}{2}$. Using now  Eq.(\ref{Eq005}) we obtain $\mathop {\lim }\limits_{\left\| {\bm{\xi }} \right\| \to \infty } {{\bm{v'}}_0} = \mathop {\lim }\limits_{\left\| {\bm{\xi }} \right\| \to \infty } \frac{{\xi _e^3}}{{2{{\left\| {\bm{\xi }} \right\|}^3}}}\left( {3\left\langle {{\bm{v}},{\bm{\hat n}}} \right\rangle {\bm{\hat n}} - {\bm{v}}} \right) - {\bm{v}} = {{\bm{v}}^\infty } = - {\bm{v}}$ which  means that
\begin{equation}\label{Eq008}
f\left( t \right) = \frac{{{{\left\| {\bm{v}} \right\|}^2}}}{2}.	
\end{equation}
Inserting Eq.(\ref{Eq005}) and Eq.(\ref{Eq006}) in Eq.(\ref{Eq007}), after some  algebraic manipulations,
we obtain the following equation:
\begin{align}\label{Eq009}
&\frac{{1}}{2} {{\left\| {\frac{3}{2}\left( {\left\langle {{\bm{v}},{\bm{\hat n}}} \right\rangle {\bm{\hat n}} - {\bm{v}}} \right) + {{\dot \xi }_e}{\bm{\hat n}}} \right\|}^2} - \frac{{{{\left\| {\bm{v}} \right\|}^2}}}{2}  - \frac{3}{2}{{\dot \xi }_e}\left\langle {{\bm{v}},{\bm{\hat n}}} \right\rangle \nonumber\\
 &+ \left( {\frac{P}{\rho } + {\Phi _{\bm{g}}}} \right)  - \frac{1}{2}\left\langle {{\bm{A}},{\bm{\hat n}}} \right\rangle  - {{\ddot \xi }_e}{\xi _e} - \frac{3}{2}\dot \xi _e^2 = 0.
\end{align}
To simplify further this equation we exploit spherical coordinates, motivated by the assumption of spherical symmetry made  for the convective elements and  the whole star (which retains its spherical shape during its existence). With the aid of this, at the surface of a convective element we write
\begin{eqnarray}
\frac{{\partial \Phi '}}{{\partial t}} &=&  - \frac{3}{2}A{\xi _e}\cos \phi  - \frac{3}{2}{{\dot \xi }_e}v\cos \theta  - {{\ddot \xi }_e}{\xi _e} - 2{{\dot \xi }_e}^2, \nonumber\\
\frac{{{{\left\| {{{{\bm{v'}}}_0}} \right\|}^2}}}{2} &=& \frac{1}{2}  \left( {\dot \xi _e^2} + \frac{9}{4}{v^2}{\sin ^2}\theta \right).  \nonumber
\end{eqnarray}
Finally by including these equations in Eq. \eqref{Eq009} we complete the proof of Eq.(\ref{Eq010}) (Q.E.D.).

This is a rather complex PDE that links the fundamental quantities to which the generic convective element is subjected within  the plasma inside the star. Nevertheless, despite its correctness, this equation is practically useless in this form because of its complexity. It is numerically solvable, but the lack of initial conditions to constrain the motion of the element does not allow us a complete coverage of the parameter space for the LHS (left-hand side) of Eq.(\ref{Eq010}). Nevertheless, it is the cornerstone of the new theory we are proposing. In order to achieve a deeper insight into its physical meaning we need to proceed with a further assumption.

\subsection{The velocity-space expansion factor}
As a convective element expands during the upward motion, its surface  acts as a piston compressing the surrounding medium and the perturbation rapidly reaches the sound speed, $v_s$ \citep[e.g.][]{1959flme.book.....L}. Under the approximations made for Eq.(\ref{Eq001}) and Eq.(\ref{Eq003}), i.e. excluding  attenuation by shear, bulk or relaxation viscosity, neglecting for the moment the heat conductivity, and limiting ourselves to the case of convective elements moving with velocities smaller than the sound speed, we obtain the following condition
\begin{equation}\label{Eq011}
	  {\varepsilon  \equiv \frac{v}{{{{\dot \xi }_e}}} \ll 1}{\forall t}> {\hat t},
\end{equation}
i.e. the relative velocity between the convective element and the intra-stellar medium $v = \left\| {\bm{v}} \right\|$ is much smaller than its expansion velocity $\dot \xi_e  = \left\| {\bm{\dot \xi}_e } \right\|$. This is a reasonable assumption for the stars and phases that we want to consider. A simple, largely intuitive justification of Eq. (\ref{Eq011}) is provided by the following arguments: an ascending  bubble must first counteract the gravity and push the surrounding medium, this second effect occurring at nearly constant gravity; therefore $v \ll \dot{\xi}_e$. In contrast, a descending bubble is accelerated by the gravity while being squeezed by the surrounding medium at nearly constant gravity and therefore its radius shrinks  faster than the descending motion, also in this case $v \ll \dot{\xi}_{e}$. Trans/supersonic motions of the convective cells (e.g. expected in red supergiants), $v \sim {v_s}$, require a fully compressive model that is beyond the aims of the present paper. We will show in Section \ref{Results} that the theory developed under the approximation of Eq.(\ref{Eq011}) leads to correct predictions for the properties of the Sun.

\textbf{Lemma 1: Pressure-radius relation for rapidly expanding/contracting sphere in an external environment}.
We prove that in the case that a sphere is expanding/contracting more rapidly than its translational motion, then the following approximate relation holds:
\begin{equation}\label{Eq012}
 \frac{P}{\rho } + {\Phi _{\bm{g}}} =  A{\xi _e}\left( {\frac{3}{2}\cos \theta  - \cos \phi } \right) + {{\ddot \xi }_e}{\xi _e} + \frac{3}{2}\dot \xi _e^2,
\end{equation}
where pressure, density and potential are evaluated at the convective element surface.

\textbf{Proof}: We start by considering the result of the previous Theorem in the form expressed by Eq.(\ref{Eq009}). We are assuming that condition expressed by Eq. (\ref{Eq011}) holds in the same environment where the Theorem is considered. Dividing both sides of Eq.(\ref{Eq009}) by $\dot \xi _e^2$ when
${\dot \xi _e} \ne 0$, i.e. formally when $\hat t \ne 0$, we can find a time ${\hat t}$ so that for $t > {\hat t} \ne 0$ we have
\begin{equation}
	{\left( {\frac{v}{{{\dot \xi _e}}}} \right)^2}\frac{1}{2}\left( {\frac{9}{4}{{\sin }^2}\theta  - 1} \right) \ll {A\frac{{{\xi _e}}}{{\dot \xi _e^2}}\left( {\frac{3}{2}\cos \theta  - \cos \phi } \right) + \frac{{{{\ddot \xi }_e}{\xi _e}}}{{\dot \xi _e^2}}}, \nonumber
\end{equation}
and
\begin{equation}
	{\left( {\frac{{v{{\dot \xi }_e}}}{{\dot \xi _e^2}}} \right)^2}\frac{5}{2}\cos \theta  \ll {A\frac{{{\xi _e}}}{{\dot \xi _e^2}}\left( {\frac{3}{2}\cos \theta  - \cos \phi } \right) + \frac{{{{\ddot \xi }_e}{\xi _e}}}{{\dot \xi _e^2}}},\nonumber
\end{equation}
thus yielding
\begin{align}
	\frac{P}{\rho } + {\Phi _{\bm{g}}} &= {\frac{A}{2}\left\| {\bm{\xi }} \right\|\frac{{\xi _e^3}}{{{{\left\| {\bm{\xi }} \right\|}^3}}}\cos \theta  + \frac{{{\xi _e}}}{{\left\| {\bm{\xi }} \right\|}}\left( {{{\ddot \xi }_e}{\xi _e} + 2\dot \xi _e^2} \right)}\nonumber\\
	&-{\left. {  \frac{1}{2}\dot \xi _e^2{{\left( {\frac{{\xi _e^2}}{{{{\left\| {\bm{\xi }} \right\|}^2}}}} \right)}^2}} \right|_{\left\| {\bm{\xi }} \right\| = {\xi _e}}},
\end{align}
where pressure, density and potential are evaluated at the convective element surface, and with ${\hat t}$ we do not refer to any ``initial time'' for the existence of a generic convective cell when ${\dot \xi _e} \sim v$, but rather to any time at which the Eq.(\ref{Eq011}) is fully satisfied. Simplifying and exploiting spherical coordinates we obtain Eq.\eqref{Eq012}. This proves the Lemma 1 (Q.E.D.).

Initially the expansion rate is not necessarily faster than the bubble speed and Eq.(\ref{Eq011}) is satisfied only asymptotically for $t$ larger than a given ${\hat t}$ that can depend on the stellar properties. The acceleration term $A$ has to be retained because the condition Eq.(\ref{Eq011}) can relate our two Lagrangian variables only by integration/derivation, but we have not yet obtained this relation as a function of the Lagrangian variables. This prevents us from performing an integration or derivation of Eq.(\ref{Eq011}) being not yet explicit the relation between acceleration and velocity: we will see only in Section \ref{Accele} that indeed it is $A = A\left( {{\bf{x}},{\bf{v}},{{{\bf{\dot \xi }}}_e}} \right)$, and the correct relation between acceleration, position, motion and expansion will be worked out only in Eq.(\ref{Eq068}) in relation with the radiative and adiabatic gradients.

We move now to a realistic situation. We consider the case  in which a convective  element moves radially upward throughout the external zones of a star and  the
acceleration and velocity are co-linear, and finally we exclude the possibility of convective overshooting that will be considered in a forthcoming paper  \citep[][]{Pasetto14}. Since the  mathematical simplification of Eq.(\ref{Eq009})  brought by   Eq.(\ref{Eq011})  is of paramount importance, we must fully understand its physical implication and meaning of it.
This theory of  convection is based on the assumption of non-local-equilibrium, i.e. we assume that the interstellar plasma on the surface of the expanding/contracting convective element while moving outward/inward slightly deviates from strict hydrostatic equilibrium.  The condition of rigorous hydrostatic equilibrium is met  by the star only at larger distances from the surface of a convective element, as already done for Eq.(\ref{Eq008}).
In this way, the property of hydrostatic equilibrium to which we refer by pushing to infinity the limit ${\bm{x}} \to \infty$ in Eq.(\ref{Eq007}) in order to fix Eq.(\ref{Eq008}), has the physical meaning of ``far away'' from the convective element surface, but still ``close enough'' to retain the density $\rho$ as constant on global stellar scale. We refer to these mathematically asymptotic but local values for the pressure as ${{P^\infty }}$, with  ${{\rho^\infty }}$ and $\Phi^\infty _{\bm{g}}$ as already mentioned in Eq.\eqref{Eq000}. The corresponding equation of hydrostatic equilibrium reads
\begin{align}\label{Eq013}
{\left. {\frac{{{\nabla _{\bm{x}}}P\left( {\bm{\xi }} \right)}}{{\rho \left( {\bm{\xi }} \right)}}} \right|_{\left\| {\bm{\xi }} \right\| \to \infty }} &= {\left. {{\bm{g}}\left( {\bm{\xi }} \right)} \right|_{\left\| {\bm{\xi }} \right\| \to \infty }}  \nonumber\\
{\left. {\frac{{{\nabla _{\bm{x}}}P\left( {\bm{\xi }} \right)}}{{\rho \left( {\bm{\xi }} \right)}}} \right|_{\left\| {\bm{\xi }} \right\| \to \infty }} &= {\left. { - \nabla {\Phi _{\bm{g}}}\left( {\bm{\xi }} \right)} \right|_{\left\| {\bm{\xi }} \right\| \to \infty }}  \nonumber\\
\frac{{{P^\infty }}}{\rho^\infty } + {\Phi^\infty _{\bm{g}}} &= 0
\end{align}
where the third equation holds by integration of the second at equilibrium (i.e. $\frac{\partial }{{\partial t}} = 0$) \citep[see for instance][]{2004cgps.book.....W,2013sse..book.....K}

We consider now the unlikely situation in which the convective element moves outward travelling through the entire star  preserving its identity until it reaches the outer layers of the star. In the co-moving reference ${S_1}$ the element surface expands until it reaches the equilibrium with the surrounding medium (note that this situation is also in strong contradiction with the standard formulation of the MLT). Thus the element reaches the kinetic limit $v \gg {\dot \xi _e}$ opposite to that considered in Eq.(\ref{Eq012}), i.e. the element surface no longer expands and in $S_1$ is in static equilibrium  (or in ${S_0}$ the element rises with constant ${\xi _e}$). In this case, the element is able to travel long distances keeping its size unchanged (apart from an initial phase of oscillations at the surface  not to be mistaken  with the Brunt-Vaisala oscillations of the element position in the layers of a star stable against  convection)\footnote{Note that in such a  case it might be necessary to include  the surface tension by means of  the Young-Laplace equation that must be  included in the EoM. This is not the limit of interest for us.}. This situation does not apply here because it is is ruled out by the conditions of Eq.(\ref{Eq011}).

We now call ${A^\infty } = A\left( {{{\bm{x}}^\infty };t} \right)$ the direction-dependent relative acceleration between $S_1$ and $S_0$ to the same limit where Eq.(\ref{Eq011}) holds and we omit now on the explicit dependence on the velocity-space. The behaviour of this term is complicated and requires a careful treatment for which we reserve all of Section \ref{Accele}. We \textsl{assume} here that this term is approximately constant for the physical system under consideration and we will provide a rigorous proof of this assumption in Section \ref{Accele}. Under this hypothesis we prove the following:

\textbf{Corollary 1: the asymptotic expansion equation for the convective element.}
In a stellar layer where ${A^\infty }\cong \rm{const.}$, the expansion of the convective element is governed asymptotically in the time evolution by the following equation:
\begin{equation}\label{Eq018}
	{{{\ddot \xi }_e}{\xi _e} + \frac{3}{2}\dot \xi _e^2 + \frac{{A^\infty{\xi _e}}}{2} = 0},
\end{equation}
\textbf{Proof}: When considering Eq.(\ref{Eq013}) it is simple to prove that the LHS of Eq.(\ref{Eq012}) cancels: $\frac{\rho }{{{P^\infty }}}\left( {\frac{P}{\rho } + {\Phi _{\bm{g}}}} \right) = \frac{\rho }{{{P^\infty }}}\left( {\frac{{P + \rho {\Phi _{\bm{g}}}}}{\rho }} \right) = \frac{\rho }{{{P^\infty }}}\left( {\frac{{P - {P^\infty }}}{\rho }} \right) = \frac{P}{{{P^\infty }}} - 1$ which goes to zero as $t > \hat t$ and $\left\| {\bm{\xi }} \right\| \to \infty $ because $P \to {P^\infty }$ independently from any angular dependence. Hence Theorem Eq.\eqref{Eq010} with Lemma 1 and this consideration results in the corollary Eq.\eqref{Eq018} and we conclude (Q.E.D.).

The equation of this corollary governs the temporal asymptotic behaviour of the convective element. Its solution is a difficult task achieved in the next section.

\subsection{Solution of the equation for a convective element in ${S_1}$}\label{EqCnv}
As Eq. (\ref{Eq018}) governs the asymptotic evolution of any convective element, it is important to cast it in a dimensionless form  and derive its most general solution.
Even though  Eq.(\ref{Eq018})   looks relatively simple, actually it is not, because of its high non-linearity. Indeed it contains two non-linear terms for the dependent variable, ${\ddot \xi _e}{\xi _e}$ and $\frac{3}{2}\dot \xi _e^2$,  and  must be coupled with another differential equation for the acceleration   ${\bm{A}^\infty}$ to form a system of two coupled PDEs. To cope with this difficulty, we start recasting  Eq.(\ref{Eq018}) by means of  dimensionless  variables.
\begin{equation}\label{Eq019}
\begin{array}{*{20}{c}}
	{\chi  \equiv \frac{{{\xi _e}}}{{{\xi _0}}}}& \rm{and} &{\tau  \equiv \frac{t}{{{t_0}}}}
	\end{array},
\end{equation}
so that for any given initial size ${\xi _0}$ of a convective element  at the initial time $t = 0$ in units of $t_0$ we have

\begin{equation}\label{Eq020}
\begin{array}{*{20}{c}}
	{\chi \left( 0 \right) = 1}& \rm{and} &{\frac{{d\chi \left( 0 \right)}}{{d\tau }} = 0}
	\end{array},
\end{equation}
according to which we have assumed that a generic convective element of  any arbitrary size starts expanding with zero expansion velocity.  We remark that this choice for the initial conditions is arbitrary. As we are interested in the asymptotic behaviour of the solution (for $\tau \gg \hat \tau$ with $\hat \tau \equiv \frac{\hat t}{t_0}$), any other initial conditions, such as $\chi(\hat \tau) = \chi_0$ and ${\frac{{d\chi \left( \hat \tau \right)}}{{d\tau }} > 0}$, would yield the same results. Therefore  $\hat \tau$ can be chosen arbitrarily close to ``0'' and considered as a dimension-less parameter.
With these assumptions we rewrite Eq.(\ref{Eq018}) as

\begin{equation}\label{Eq021}
\chi \frac{{{d^2}\chi }}{{d{\tau ^2}}} + \frac{3}{2}{\left( {\frac{{d\chi }}{{d\tau }}} \right)^2} + \frac{A^\infty}{2}\frac{{t_0^2}}{{{\xi _0}}}\chi  = 0.
\end{equation}
In this equation the normalized acceleration  is a function of the time and position, the dependencies of which will be investigated in detail in  Section \ref{Accele} below. In the previous Section we have seen the relation between the condition Eq.(\ref{Eq011}) and the reduced spatial motion travelled by a convective element. Here we assumed that:
\begin{equation}\label{Eq022}
		A^\infty\left( {\chi;\tau  } \right) \cong \rm{const.}
\end{equation}
to solve Eq.(\ref{Eq021}), deferring a rigorous proof of this assumption to a devoted corollary in the next section.
At this point one could try to find a numerical solution of the equations as  functions of time and space provided the temporal and spatial evolution of the acceleration is known. However, this way of proceeding would not improve significantly the theory of convection. This goal can be achieved
by pushing the analytical analysis of the problem further. We continue  to Eq.(\ref{Eq021}) in fully non-dimensional form by assuming
\begin{equation}\label{Eq015}
\frac{A^\infty}{2}\frac{{t_0^2}}{{{\xi _0}}} \equiv \frac{1}{2}\frac{A^\infty}{{{A^\infty_0}}} \equiv  - 2.
\end{equation}
The reason for the last equality to $-2$ will become clear later on:
it simply allows us to account for the fact that in $S_1$ the acceleration
of the convective element is due only  to the surface expansion
and to the opposite motion of the intra-stellar fluid on the
surface of the convective element itself. The factor 2 is introduced
for mathematical convenience. Now
\begin{equation}\label{Eq024}
\chi \frac{{{d^2}\chi }}{{d{\tau ^2}}} + \frac{3}{2}{\left( {\frac{{d\chi }}{{d\tau }}} \right)^2} - 2\chi  = 0,
\end{equation}
whose solution is obtained in Appendix A (being simply a mathematical problem). The $\chi$'s asymptotic solution of interest for $t \to \infty $ (the interested reader can look at Appendix A) is:
\begin{eqnarray}\label{Eq031}
\chi \left( \tau  \right) &=& \frac{1}{{{8^2}}}\frac{{{{\left( {\sqrt \pi  \Gamma \left( { - {\raise0.5ex\hbox{$\scriptstyle 1$}
\kern-0.1em/\kern-0.15em
\lower0.25ex\hbox{$\scriptstyle 8$}}} \right) - \Gamma \left( {{\raise0.5ex\hbox{$\scriptstyle 3$}
\kern-0.1em/\kern-0.15em
\lower0.25ex\hbox{$\scriptstyle 8$}}} \right)\tau } \right)}^2}}}{{\Gamma {{\left( {{\raise0.5ex\hbox{$\scriptstyle 3$}
\kern-0.1em/\kern-0.15em
\lower0.25ex\hbox{$\scriptstyle 8$}}} \right)}^2}}} + \textrm{O}{\left( \chi  \right)^3} \nonumber\\
 &\simeq& \frac{1}{{{8^2}}}{\tau ^2} - \frac{{\sqrt \pi  \Gamma \left( { - {\raise0.5ex\hbox{$\scriptstyle 1$}
\kern-0.1em/\kern-0.15em
\lower0.25ex\hbox{$\scriptstyle 8$}}} \right)}}{{32\Gamma \left( {{\raise0.5ex\hbox{$\scriptstyle 3$}
\kern-0.1em/\kern-0.15em
\lower0.25ex\hbox{$\scriptstyle 8$}}} \right)}}\tau  + \frac{{\pi \Gamma {{\left( { - {\raise0.5ex\hbox{$\scriptstyle 1$}
\kern-0.1em/\kern-0.15em
\lower0.25ex\hbox{$\scriptstyle 8$}}} \right)}^2}}}{{64\Gamma {{\left( {{\raise0.5ex\hbox{$\scriptstyle 3$}
\kern-0.1em/\kern-0.15em
\lower0.25ex\hbox{$\scriptstyle 8$}}} \right)}^2}}}.
\end{eqnarray}
The $\chi$ asymptotic dependence is $ \sim {\tau ^2}$ plus lower order correction terms, i.e. quadratic in $\tau$ for $t \to \infty $ (see our cautionary remark made at the beginning of the analysis at the bottom of Section \ref{probIntro2}). As a consequence of this, also the time averaged value $\bar \chi \left( \tau  \right) = \frac{1}{\tau }\int_0^\tau  {\chi \left( {\tau '} \right)d\tau '}$ will grow with the same temporal dependence. This is an extremely important result that will play a key role in the proof of the independence of the stellar convection from any free parameter (see Section \ref{EqSc} below).
These results fully determine the relationship between the motion and  the expansion (contraction) of a convective element.

Finally, it is easy to prove that once we are interested in the integration of a star in a phase of non-hydrostatic equilibrium, i.e. where $\frac{{{P^\infty }}}{{{\rho ^\infty }}} + \Phi _{\bm{g}}^\infty \neq 0$, equation Eq.(\ref{Eq012}) is again integrated numerically with $\frac{{P\left( {{{\bm{x}}}} \right)}}{\rho } + {\Phi _{\bm{g}}}\left( {{{\bm{x}}}} \right)$ being a known term valid everywhere in the stellar model. Thus, the solution to Eq.(\ref{Eq012}) can equally be recovered by simple translation of the solution for $\chi$ presented above, e.g. with a re-normalization to ${\raise0.5ex\hbox{$\scriptstyle {\frac{{{P^\infty }}}{{{\rho ^\infty }}} - \Phi _{\bm{g}}^\infty }$}
\kern-0.1em/\kern-0.15em
\lower0.25ex\hbox{$\scriptstyle {{A^\infty }}$}} \equiv  - 2$ being $A^\infty$ bounded in spherical coordinates $\left( {\frac{3}{2}\cos \theta  - \cos \phi } \right) \in \left[ { - \frac{5}{2},\frac{5}{2}} \right]$.

With this approximation, we are simplifying the system of two coupled PDEs, one for the expansion radius ${\xi _e}$ and one for the acceleration  ${\bm{A}^\infty} = {{\bm{\ddot x}}_{O'}}$ of a convective element,  to  a  new system of two coupled  DEs; these being Eq.(\ref{Eq024}) and an equation for the acceleration to be developed now. The next issue to address is therefore now to prove a corollary on the acceleration in Eq.(\ref{Eq022}). This is the subject of the Section \ref{Accele} below.

\section{The acceleration of a convective element}\label{Accele}
So fare we have dealt with the first degree of freedom, the convective element radius $\xi_e$. In this section we examine the forces acting on the convective element and ruling its motion, i.e. we deal with the second degree of freedom of our system, the motion of the convective element initially located at a generic position inside the star, ${{\bf{x}}_{O'}}$.
The PDE Eq.(\ref{Eq010}) or its approximated ODE Eq.(\ref{Eq018}) has to be supplied with a second ODE describing the motion of a convective element throughout the stellar medium.  This yields  a system of two equations with two unknowns, i.e. ${\xi _e} = {\xi _e}\left( t \right)$ and ${{\bm{x}}_{O'}} = {{\bm{x}}_{O'}}\left( t \right)$. We present the detailed derivation of these equations from basic principles in order to explain the precise founding hypotheses of these  equations. The arguments that lead to this result hold only in the context of the Theorem of Section \ref{EqcnvQ}. In the same hypothesis of this theorem the following corollary holds:

\textbf{Corollary 2: Acceleration of the convective element.}
In a stellar layer under the assumption of Theorem Eq.\eqref{Eq010} the acceleration of a convective element is given by
\begin{equation}\label{Eq033}
{\bm{A}} =  - {\bm{g}}\frac{{{m_e} - M}}{{{m_e} + \frac{M}{2}}} - \frac{{10}}{3}\pi \xi _e^2\rho {\bm{v}}{\dot \xi _e}.
\end{equation}
\textbf{Proof}: The total force acting on a convective element in ${S_1}$ is determined by the total pressure acting on its surface ${{\bm{F}}_P} =  - \int_{}^{} {P{\bm{\hat n}}d\Omega } $ and  the weight  ${m_e}{\bm{g}}$ of the element. The corresponding  equation of motion  derived from the Newtonian law ${\bm{F}} = m{\bm{A}} = m{\bm{\dot v}} = m{{\bm{\ddot x}}_{O'}}$ is the equation to be integrated  together with Eq.(\ref{Eq010}).
In the reference frame  ${S_1}$ we can use this force balance to express the pressure force ${{\bm{F}}_P}$  and in turn relate it to the acceleration.
We may write
\begin{eqnarray}\label{Eq032}
 - \int_{}^{} {P{\bm{\hat n}}d\Omega }  &=& - \rho \int_{}^{} {\left( {\frac{{{v^2}}}{2}\left( {1 - \frac{9}{4}{{\sin }^2}\theta } \right) + \frac{5}{2}v{{\dot \xi }_e}\cos \theta } \right.}  \nonumber\\
																				&+& \left. { \frac{{A{\xi _e}}}{2}\cos \theta  + {{\ddot \xi }_e}{\xi _e} + \frac{3}{2}\dot \xi _e^2 - {\Phi _{\bm{g}}}} \right){\bm{\hat n}}d\Omega \nonumber\\
																				&=& - \rho \int_{}^{} {\left( {\frac{{{v^2}}}{2}\left( {1 - \frac{9}{4}{{\sin }^2}\theta } \right) + \frac{5}{2}v{{\dot \xi }_e}\cos \theta } \right.}  \nonumber\\
																				&+& \left. { \frac{{A{\xi _e}}}{2}\cos \theta  + {{\ddot \xi }_e}{\xi _e} + \frac{3}{2}\dot \xi _e^2 - {\Phi _{\bm{g}}}} \right)\mathcal{J}  {\bm{\hat n}}d\Omega   \nonumber\\
																				&+& \rho \int_{}^{} {\nabla {\Phi _{\bm{g}}}{d^3}{\xi _e}} \nonumber\\
																				&=&  - \frac{2}{3}\pi \xi _e^3\rho {\bm{A}} - \frac{{10}}{3}\pi \xi _e^2\rho {\bm{v}}{\dot \xi _e} + \frac{4}{3}\pi \xi _e^3\rho {\bm{g}},
\end{eqnarray}
where the integral is carried out over  the sphere representing the convective element at each instant on the differential form (the solid angle) $d\Omega$. $\theta$ is the angle already defined after Eq.(\ref{Eq009}),  $\mathcal{J}$ is the Jacobian of the transformation from cartesian to spherical coordinates, and ${\bm{\hat n}}$ is the unit vector of the position vector. Finally in the right hand side (RHS) of the equation, third line, simple trigonometric integrals have been computed and the Gauss theorem has been used to worked out explicitly the result. This equation accounts for the buoyancy of the convective element $\frac{4}{3}\pi \xi _e^3\rho {\bm{g}}$,  the inertial term of the fluid  displaced by the movement of the convective cell, i.e. the reaction mass $\frac{1}{2}\frac{4}{3}\pi \xi _e^3\rho  \equiv \frac{M}{2}$, and a new extra term $ - \frac{{10}}{3}\pi \xi _e^2\rho {\bm{v}}{\dot \xi _e}$ arising from the changing  size of the convective element: the larger  the convective element, the stronger  the buoyancy effect and the larger is the velocity acquired by  the convective element. These terms have to be included in the Newtonian  EoM.  Adding now  the effect of the gravity  on the convective element of mass ${m_e}$ we get the general expression for the acceleration {\bf A} in $S_1$ as in Eq.\eqref{Eq033} (Q.E.D.).

It is worth calling attention to the correction to the force balance that is required  in order to  properly include  the inertia that  a convective element experiences during its motion across the stellar fluid. More precisely,  a convective element in  motion experiences a drag force produced by the different density of the fluid it is moving through. In this way we  naturally reconcile the correct physics with the D'Alambert paradox intrinsic to  the velocity-potential theory approximation.  If in Eq.(\ref{Eq033}) we change sign to recast it  in ${S_0}$ instead of ${S_1}$ we recover standard results from fluid dynamics for the force balance, e.g. Eq.(6.8.20) of  \citet{2000ifd..book.....B}.  Examining Eq.(\ref{Eq033}) we note that  for ${m_e} = M$ we have ${\bm{A}} = 0$ as indeed expected if there is no overdensity (no convection). For ${m_e} \ll M$,  ${\bm{A}} \simeq  - 2{\bm{g}}$, which means that the convective element is reaching  a limiting acceleration, i.e. the case excluded in the convection regime. For ${m_e} \gg M$ the fluid scarcely affects the initial acceleration of a convective element.
This apparently means that the approximation in Eq.(\ref{Eq022}) does not hold, because the expansion rate of the bubble ${\dot \xi _e} \ne 0$ as clearly assumed in the previous section and so apparently $A \ne 0$. To show that this is not the case is simple by taking into account again Eq.(\ref{Eq011}) and we examine this as a case of interest.

\textit{The case of interest: rapidly expanding convective element}. As simple application of this corollary we look at the case of a convective element rising along the vertical direction in the reference frame ${S_0}$ with ${\bm{g}} = \left\{ {0,0, - g} \right\}$ and $g > 0$, and considering the same  approximation used for Eq.(\ref{Eq011}) and notation used for Eq.\eqref{Eq013} and \eqref{Eq018} we get:
\begin{align}\label{Eq034}
	{m_e}{A_z^{\infty }} &=  - \frac{2}{3}\pi \xi _e^3\rho {A_z^{\infty }} - \frac{{10}}{3}\pi \xi _e^2\rho v{{\dot \xi }_e} - \frac{4}{3}\pi \xi _e^3\rho g + {m_e}g \Leftrightarrow \nonumber\\
	     {A_z^{\infty }} &= g\frac{{{m_e} - M}}{{{m_e} + \frac{M}{2}}} - \frac{{10}}{3}\pi \xi _e^2\rho v{{\dot \xi }_e} \nonumber\\
	           &\cong g\frac{{{m_e} - M}}{{{m_e} + \frac{M}{2}}},
\end{align}
where for ${\dot \xi _e} \ne 0$ we divide and multiply by $\dot \xi _e^2$ as already done for Eq.(\ref{Eq012}) to eliminate the term $\frac{{v{{\dot \xi }_e}}}{{\dot \xi _e^2}}$ and to formulate Eq.(\ref{Eq034}) as an asymptotic expansion of order $\textrm{O}\left( {\frac{{{A_z^{\infty }}}}{g}} \right)$. This simple exercise proves that at the same degree of approximation under which Eq.(\ref{Eq021}) holds also Eq.(\ref{Eq022}) holds - as was left to prove.
We see also that the convective element will rise when ${m_e} < M \Rightarrow {m_e} - M < 0$, i.e. when $A$ in ${S_1}$ is negative so that the sign adopted in Eq.(\ref{Eq015}) remains fully justified as originally adopted.

The last step required to integrate the EoM for a convective element within a convective layer and it deals with the instability conditions. The following auxiliary lemma proves a self-standing result that once included in the previous corollary will allow us to mathematically close the set of equations and to conclude the theory.

\textbf{Lemma 2: Linear response of the convective element to the stellar pressure gradients}.
We prove that the response, i.e. the motion, of the convective element to the forces applied on it, i.e. Eq.\eqref{Eq034}, is given in linear regime by
\begin{equation}\label{Eq043}
{A_z^{\infty }} \simeq g\frac{{{\nabla _e} - \nabla  + \frac{\varphi }{\delta }{\nabla _\mu }}}{{\frac{{3{h_P}}}{{2\delta \Delta z}} + \left( {{\nabla _e} + 2\nabla  - \frac{\varphi }{{2\delta }}{\nabla _\mu }} \right)}}.
\end{equation}
to the leading order on $\textrm{O}\left( {\frac{{\Delta {\bm{x}}}}{{{h_P}}}} \right)$.

\textbf{Proof}: We need to express the masses term in Eq.(\ref{Eq034}) as a function of  the fundamental logarithmic gradients introduced in Section \ref{Introduction}. This step  is indeed  necessary owing to the presence of the new term ${m_e} + \frac{M}{2}$ in the denominator of Eq.(\ref{Eq034}). For a small displacement of the convective element, say $\Delta {{\bm{x}}_{O'}}$ the density can be expanded as 
\begin{eqnarray}\label{Eq035}
	\rho  &=& {\left. \rho  \right|_{{{\bm{x}}_{O'}}}} + {\nabla _{\bm{x}}}{\left. \rho  \right|_{{{\bm{x}}_{O'}}}}\Delta {{\bm{x}}_{O'}} + ... \nonumber\\
	{\rho _e} &=& {\left. {{\rho _e}} \right|_{{{\bm{x}}_{O'}}}} + {\nabla _{\bm{x}}}{\left. {{\rho _e}} \right|_{{{\bm{x}}_{O'}}}}\Delta {{\bm{x}}_{O'}} + ...,
	\end{eqnarray}
in which all terms of quadratic order in $\Delta {{\bm{x}}_{O'}}$ and higher orders are neglected. The subscript 0 refers to the equilibrium position of the convective cell. Because the volume occupied by the convective element is the same displaced in the fluid, the relation Eq.(\ref{Eq034}) can easily be  translated to an equivalent one in the density. In its numerator, assuming that ${\left[ {{\rho _e} - \rho } \right]_{{{\bm{x}}_{O'}}}} = 0$,  we get the standard approximation \citep[e.g.][]{ 1994sse..book.....K}  to the first order ${\rho _e} - \rho  \simeq {\left[ {{\nabla _{\bm{x}}}{\rho _e} - {\nabla _{\bm{x}}}{\rho _e}} \right]_{{{\bm{x}}_{O'}}}}\Delta {{\bm{x}}_{O'}}$ for the displacement of a convective element. However, because of the terms  at the denominator we get the more complicated expression:
\begin{equation}\label{Eq036}
{\rho _e} + \frac{\rho }{2} \simeq \frac{3}{2}{\left. \rho  \right|_{{{\bm{x}}_{O'}}}}+ {\left[ {{\nabla _{\bm{x}}}{\rho _e} + \frac{{{\nabla _{\bm{x}}}\rho }}{2}} \right]_{{{\bm{x}}_{O'}}}}\Delta {{\bm{x}}_{O'}}.
\end{equation}
This equation requires the density gradients which do not appear naturally in the equations of stellar structure. Therefore, we express them  as a function of temperature $T$ with the help of the EoS. Now, we must recast the correspondent instability criteria starting from the EoS for a perfect fluid  $\rho  = \rho \left( {P,T,\mu } \right)$ (see Section \ref{Introduction}) that in its differential form reads $\frac{{d\rho }}{\rho } = \alpha \frac{{dP}}{P} + \delta \frac{{dT}}{T} + \varphi \frac{{d\mu }}{\mu }$. Here $\left\{ {\alpha ,\delta ,\varphi } \right\} \equiv \left\{ {\frac{{\partial \ln \rho }}{{\partial \ln P}}, - \frac{{\partial \ln \rho }}{{\partial \ln T}},\frac{{\partial \ln \rho }}{{\partial \ln \mu }}} \right\}$ where the standard notation has been used. Thus, we get
\begin{eqnarray}\label{Eq037}
{\nabla _{\bm{x}}}{\rho _e} + \frac{{{\nabla _{\bm{x}}}\rho }}{2} &=& {\left[ {\frac{{\rho \alpha }}{P}{\nabla _{\bm{x}}}P - \frac{{\rho \delta }}{T}{\nabla _{\bm{x}}}T + \frac{{\rho \varphi }}{\mu }{\nabla _{\bm{x}}}\mu } \right]_e} \nonumber\\
 &+& \frac{1}{2}\left( {\frac{{\rho \alpha }}{P}{\nabla _{\bm{x}}}P - \frac{{\rho \delta }}{T}{\nabla _{\bm{x}}}T + \frac{{\rho \varphi }}{\mu }{\nabla _{\bm{x}}}\mu } \right).
\end{eqnarray}
After some manipulation, and assuming that in a small displacement the change of molecular weight $\mu $, $d\mu  = 0$ for the moving element, we are able to simplify the RHS as
\begin{eqnarray}\label{Eq038}
 {\rm{RHS}}&=& {\rho _e}{\left[ {\frac{\alpha }{P}{\nabla _{\bm{x}}}P} \right]_e} - {\rho _e}{\left[ {\frac{\delta }{T}{\nabla _{\bm{x}}}T} \right]_e}  \nonumber\\
&+& \frac{\rho }{2}\frac{\alpha }{P}{\nabla _{\bm{x}}}P - \frac{\rho }{2}\frac{\delta }{T}{\nabla _{\bm{x}}}T + \frac{\rho }{2}\frac{\varphi }{\mu }{\nabla _{\bm{x}}}\mu.
\end{eqnarray}
We consider now the first and the third element in the previous Eq.\eqref{Eq038}. While in the case of the derivation of the Schwarzschild criteria the first two terms in the previous equation simplify because ${P_e} - P = 0$ in any adiabatic expansion, this is no longer the case here. We have
\begin{eqnarray}\label{Eq039}
	{\rho _e}{\left[ {\frac{\alpha }{P}{\nabla _{\bm{x}}}P} \right]_e} + \frac{\rho }{2}\frac{\alpha }{P}{\nabla _{\bm{x}}}P
	&\cong& \rho \alpha \left( {\frac{1}{{{P_e}}}{\nabla _{\bm{x}}}{P_e} + \frac{1}{2}\frac{1}{P}{\nabla _{\bm{x}}}P} \right) \nonumber\\
	&=& \frac{{\rho \alpha }}{P}\left( {1 + \frac{1}{2}} \right){\nabla _{\bm{x}}}P \nonumber\\
	&=& \frac{3}{2}\frac{{\rho \alpha }}{P}{\nabla _{\bm{x}}}P,
	\end{eqnarray}
which is used to simplify Eq.(\ref{Eq038}) as:
\begin{equation}\label{Eq040}
{\rm{RHS}}=\frac{3}{2}\frac{{\rho \alpha }}{P}{\nabla _{\bm{x}}}P - \rho {\left[ {\frac{\delta }{T}{\nabla _{\bm{x}}}T} \right]_e} - \frac{\rho }{2}\frac{\delta }{T}{\nabla _{\bm{x}}}T + \frac{\rho }{2}\frac{\varphi }{\mu }{\nabla _{\bm{x}}}\mu.
\end{equation}
Now, by introducing the pressure scale length, with $\bm x = \{0,0,z\}$,  we can further expand the previous Eq.\eqref{Eq040} as:
\begin{align}\label{Eq041}
{\rm{RHS}} &=\frac{\rho }{{{h_P}}}\frac{{3\alpha }}{{2P}}\left( { - P\frac{{dz}}{{dP}}} \right)\frac{{dP}}{{dz}} - \frac{\rho }{{{h_P}}}{\left[ {\left( { - P\frac{{dz}}{{dP}}} \right)\frac{\delta }{T}\frac{{dT}}{{dz}}} \right]_e} \nonumber\\
 &- \frac{\rho }{{{h_P}}}\left( { - P\frac{{dz}}{{dP}}} \right)\frac{1}{2}\frac{\delta }{T}\frac{{dT}}{{dz}} + \frac{\rho }{{{h_P}}}\left( { - P\frac{{dz}}{{dP}}} \right)\frac{1}{2}\frac{\varphi }{\mu }\frac{{d\mu }}{{dz}}.
	\end{align}
If we introduce the logarithmic derivative to write this equation as a function of the standard temperature gradients:
\begin{align}\label{Eq042}
 {\rm{RHS}} &=  - \frac{\rho }{{{h_P}}}\frac{{3\alpha }}{2} + {\left[ {\delta \frac{{d\ln T}}{{d\ln P}}} \right]_e} + \frac{1}{2}\delta \frac{\rho }{{{h_P}}}\frac{{d\ln T}}{{d\ln P}} - \frac{\rho }{{{h_P}}}\frac{\varphi }{2}\frac{{d\ln \mu }}{{d\ln P}} \nonumber\\
 &= \frac{{\rho \delta }}{{{h_P}}}\left( { - \frac{{3\alpha }}{{2\delta }} + {\nabla _e} + \frac{1}{2}\nabla  - \frac{\varphi }{{2\delta }}{\nabla _\mu }} \right),
	\end{align}
and recalling that $\frac{\alpha }{\delta } = \frac{{{\raise0.5ex\hbox{$\scriptstyle {\partial \ln \rho }$}
\kern-0.1em/\kern-0.15em
\lower0.25ex\hbox{$\scriptstyle {\partial \ln P}$}}}}{{ - \left( {{\raise0.5ex\hbox{$\scriptstyle {\partial \ln \rho }$}
\kern-0.1em/\kern-0.15em
\lower0.25ex\hbox{$\scriptstyle {\partial \ln T}$}}} \right)}} =  - \frac{{\partial \ln T}}{{\partial \ln P}} =  - \nabla $
we can write Eq.(\ref{Eq037}) after simple algebraic manipulation as:
\[{\left. {\frac{{d\rho }}{{dz}}} \right|_e} + \frac{1}{2}\frac{{d\rho }}{{dz}} = \frac{{\rho \delta }}{{{h_P}}}\left( {{\nabla _b} + 2\nabla  - \frac{\varphi }{{2\delta }}{\nabla _\mu }} \right).\]
It is straightforward to now compute the acceleration as a function of the fundamental logarithmic gradients. By considering the motion of the convective elements along the vertical direction $z$, once we introduce this term in Eq.(\ref{Eq036}) and we consider Eq.(\ref{Eq034}) after algebraic manipulation we conclude with Eq.\eqref{Eq043} (Q.E.D.).

It is interesting now to call attention to  some aspects of the acceleration Eq.(\ref{Eq043}) with respect to the classical formulation in the literature.
\begin{enumerate}
	\item In regions of homogeneous chemical composition the acceleration  reduces to
	\begin{equation}\label{Eq044}
		{A_z^{\infty }} \simeq g\frac{{{\nabla _e} - \nabla }}{{\frac{{3{h_P}}}{{2\delta \Delta z}} + \left( {{\nabla _e} + 2\nabla } \right)}},
	\end{equation}
	It is then immediately evident how this new instability criterion induces exactly the Schwarzschild instability zones (${\nabla _e} - \nabla  < 0$) as the denominator of Eq.(\ref{Eq044}) is always positive by definition! This is a very important result  because it allows us to extend the Schwarzschild and/or Ledoux criteria for instability: even with the new criterion,  the convective zones occur exactly in the same regions predicted by the Schwarzschild  criterion.

	\item The zones of chemical inhomogeneity should be treated according to the instability criterion given by Eq. (\ref{Eq043}). In any case, we point out that  relation (\ref{Eq043}) is a second order Taylor expansion on the small parameter, $\varepsilon  = \frac{{2\delta \Delta z}}{{3{h_P}}}$, which to the first order yields the classical results for the acceleration of a convective element ${A_z^{\infty }} = g\frac{2}{3}\frac{\delta }{{{h_P}}}\left( {{\nabla _e} - \nabla  + \frac{\varphi }{\delta }{\nabla _\mu }} \right)\Delta z + \textrm{O}{\left( {\frac{{2\delta \Delta z}}{{3{h_P}}}} \right)^2}$ corrected by the factor $\frac{2}{3}$ for the presence of the inertial mass with respect to classical results. Furthermore, it   incorporates the Ledoux condition. The onset and effects of convection in the presence of a gradient in molecular weight are  highly debated subjects that are not addressed here \citep[e.g.][]{1994sse..book.....K,2009pfer.book.....M}.
\end{enumerate}
In Section \ref{EqSc} we will show that  retaining the second order terms is the key action in order to eliminate of the mixing-length thanks to Eq.(\ref{Eq010}). 

By proving  the corollary 2 and the lemma 2, we have concluded the new theory. We obtained indeed the relevant equations to govern the evolution of a convective element in an unstable convective layer inside a star. In a non-inertial reference frame $S_1$ we developed a framework where the two Eqs.\eqref{Eq018} and \eqref{Eq043} describe the evolution of the two degrees of freedom used to describe our physical system. We apply now our theory of convection in stellar interiors to a few key tests to show its potential capability.

\section{Results of the theory}\label{Results}

In this section, we present some results and predictions of our theory. We compute a few selected physical quantities of interest and then we consider their temporal evolution in the integration of a stellar model (in our case, the Sun).

Before proceeding further,  in the spirit of the cautionary remark made at the end of Section \ref{probIntro2},
we comment on the time-limits that we are going to consider. As already pointed out in Section \ref{EqCnv}, once the condition of instability to convection is matched at a certain location ${\mathbf{x}} + d{\mathbf{x}}$  inside the star, convective elements are born and initially their radius  ${\xi _e}$ and surface in turn do not necessarily expand  faster than their vertical motion $v$, i.e. initially it is $\frac{v}{{{{\dot \xi }_e}}} = O\left( 1 \right)$ as $t \to 0$.  Therefore, as the condition of Eq.\eqref{Eq011} cannot be satisfied, we must start our time integration from $t > {t_{\min }}$ (this is shown in detail in Fig. \ref{Fig2} of Appendix A).
Different arguments apply to the  upper temporal limit. Integrating the last row of Eq.\eqref{Eq034} we see that $v \propto t$ as $t \to \infty $ (this is indeed also consistent with the spatial series expansion of Eq.\eqref{Eq035} retained to the first order), but at the time  $\chi  \propto t^2$ and hence ${\xi _e} \propto t^2$ as $t \to \infty $ and finally ${\dot \xi _e} \propto t$ as $t \to \infty $. Therefore,  the condition  $\frac{v}{{{{\dot \xi }_e}}} = O\left( 1 \right)$ as $t \to \infty $ cannot be satisfied as required by Eq.\eqref{Eq011}.
To cope with this and maintain the standard notation  $\mathop {\lim }\limits_{t \to \infty } Q\left( t \right) = {Q^\infty }$ at the same time, we take a suitable time interval
$t \in \left] {{t_{\min }},{t_{\max }}} \right[$ , where at ${t_{\max }}$ the convective elements still live in an ambient medium of constant intra-stellar density (see Eq.\eqref{Eq035}) and acceleration (see Eq.\eqref{Eq022} and \eqref{Eq034}). This approach has some similarity with the Boussinesq and anelastic approximations commonly used in other branches of research such as  planetary and atmospheric sciences, oceanography, and geo-dynamics  \citep[e.g.][]{Glatzmaier2013}. The novelty here  is that for the first time all this has been formulated in the  comoving  frame of reference ${S_1}$.

\subsection{The convective flux: from the single element to the collective description}\label{convflux}
In order to apply the theory in practice, we need to consider a collective description of the convective cells. Many options (both numerical and analytical) are nowadays available to stochastically describe a phenomena within the framework of a theory, nevertheless we will limit to the simple method of the moments (MoM). The reasons are twofold: first, we will see that by limiting our analysis to the mean-stream of the convective cells (i.e. the first of a full hierarchy of moments) will leave us with extremely satisfactory results, and second, the MoM permits a more natural comparison with the MLT where only the mean velocity of convective elements is considered.

We consider an arbitrary but fixed surface $S$ inside the star with infinitesimal element $d{\bm{S}} = {\bm{\hat n}}dS$, where ${\bm{\hat n}}$ is the outward normal to the surface under consideration (i.e. any ideal surface through which the convective elements are free to flow). We assume that the number of convective elements passing through $d{\bm{S}}$ at a given time is $n = f{d^3}\bm{x}{d^3}\bm{p}$  where $f = f\left( {{\bm{x}},{\bm{p}};t} \right)$ is the unknown distribution function (DF) of the convective elements inside the star. Then for every scalar quantity of interest, say $Q$, the out/inward flux ${\varphi }$ of $Q$ is ${\varphi }\equiv\left\langle {Qf{\bm{V}},d{\bm{S}}dt{d^3}\bm{p}} \right\rangle $ with ${\bm{V}} \equiv \frac{{\bm{p}}}{{{m_e}}}$. This  represents the amount  of $Q$ transported through $d{\bm{S}}$ with a given momentum ${\bm{p}} \in \left[ {{\bm{p}},{\bm{p}} + {d^3}{\bm{p}}} \right]$ during the time interval $dt$. Therefore $\left\langle {\overline {Q{\bm{V}}} ,d{\bm{S}}} \right\rangle dt = \frac{1}{n}\int_{{{\bm{p}}^3}}^{} {\left\langle {Q{\bm{V}},d{\bm{S}}} \right\rangle fdt{d^3}\bm{p}},$ where the over-bar indicates the average of the quantity, and the amount of $Q$ transported by any convective element with any ${\bm{p}}$ through $d{\bm{S}}$ in $dt$ is $n\left\langle {\overline {Q{\bm{V}}} ,d{\bm{S}}} \right\rangle dt = \int_{{{\bm{p}}^3}}^{} {\left\langle {Q{\bm{V}},fd{\bm{S}}} \right\rangle dt{d^3}\bm{p}} $. Hence the flux of $Q$ is the amount of $Q$ per unit area and unit time\footnote{A preliminary investigation suggests that  the present formalism (see the relationship between  $\chi$ and $\xi$ of Fig. \ref{Fig2}, the connection between the size scale $\xi_0$ and the acceleration $A$, and finally the relationship between a star's gravitational stratification and $\xi_0$) could suggest the shape of the DF, $f$. However, we remind that despite only the infinite series of moments in a MoM method is mathematically equivalent to the underlying  DF, for the the sake of simplicity we prefer to adopt here the classical definition of convective flux based on the average (first order moment) velocity of convective elements. We defer a Monte Carlo approach to future studies of the DF.}
\begin{equation}\label{Eq045}
{\bm{\varphi }} = n\overline {Q{\bm{V}}}.
\end{equation}
In order to calculate the \textit{convective flux} we need to compute the amount of internal energy per unit area per unit time carried by the convective elements. In our previous computation we assumed that in asymptotic regime the convective elements move adiabatically. Recalling  that by definition the specific heat at constant pressure ${c_P}$ is the amount of heat required to increase the temperature of a convective element of unit mass by one degree, we set ${c_P} \equiv \frac{1}{{{m_e}}}\frac{{dQ}}{{dT}}$ where the pressure $P=\rm{const}.$ and ${m_e}=1$ and we have replaced the ``$Q$'' symbol of ``quantity'' defined above before with its meaning of ``heat''. Therefore,
\begin{equation}\label{Eq046}
\Delta Q \equiv {m_e}{c_P}\Delta T,
\end{equation}
is the heat excess of a convective element of mass ${m_e}$ over the surroundings. We must use $c_P$, rather than $c_V$, here, in accordance with our assumption of pressure equilibrium, the heat exchange with the surrounding medium occurs at constant pressure at each level.  Once the element has moved  from its initial position ${{\bm{x}}_1}$ with temperature ${T_{e,1}} = {T_e}\left( {{{\bm{x}}_1}} \right)$ and ambient temperature ${T_1} = T\left( {{{\bm{x}}_1}} \right)$ to a new position at distance ${{\bm{x}}_2}$ with temperature ${T_{e,2}} = {T_e}\left( {{{\bm{x}}_2}} \right)$ and ambient temperature ${T_2} = T\left( {{{\bm{x}}_2}} \right)$, the heat stored in the element  flows from this  to the surrounding medium (or vice versa depending on the ratio between ${T_{e2}}$ and ${T_2}$). The amount of heat flowing through $d{\bm{S}}$ in $dt$  is written as $Q = {c_P}\Delta T\rho \left\langle {{\bm{\bar V}},d{\bm{S}}dt} \right\rangle $ where  $\rho {\bm{\bar V}}$ is the mass flux. Consequently, the convective flux (i.e. heat passing through the surface $d{\bm{S}}$ in $dt$) is
\begin{equation}\label{Eq047}
{{\bm{\varphi }}_{\rm{cnv}}} \equiv  {c_P}\Delta T\rho {\bm{\bar V}},
\end{equation}
with ${\bm{\bar V}}$ the average velocity of the convective elements during the time interval $dt$ as seen in ${S_0}$.
Here, we see that even if there is no net mass flux, the heat is anyway transported, because this flux in ${S_0}$ can be written as the sum
\begin{eqnarray}\label{Eq048}
	{{\bm{\varphi }}_{\rm{cnv}}} &=&  n\overline {{m_e}{c_P}\Delta T{\bm{V}}} \nonumber\\	
											&=&  \rho {c_P}\overline {\Delta T{\bm{v}}}  + \rho {c_P}\overline {\Delta T{{\bm{v}}_0}} ,
\end{eqnarray}
where ${\bm{v}} = {\bm{V}} - {{\bm{v}}_0}$ is the peculiar velocity of a convective element  in Eq.(\ref{Eq004}). If the mean velocity is zero, i.e. if there is no outflow/inflow mass flux carrying the heat, the energy transport owing to the heat carried by the  convective elements 	${{\bm{\varphi }}_{\rm{cnv}}} =  \rho {c_P}\overline {\Delta T{\bm{v}}} $.

We can calculate a self-consistent  expression for the velocity  in ${S_1}$ from Eq.(\ref{Eq009}). If $\bm{v} = 0$ the heat flux carried by each  convective element is null (the convective element is supposed to be spherical) because the same stagnation point will exist in the diametrically opposite side of the convective element. Therefore, what contributes to the convective flux is not the velocity of the stagnation points, but the velocity of the whole convective element, i.e. the velocity of the barycentre.
Starting from the general expression for the velocity of a fluid  element  impacting  the surface of a convective element given by  Eq.(\ref{Eq009}), to get the sole motion of the center of the convective element it is sufficient to set
${\dot \xi _e} = {\ddot \xi _e} = 0$.  This means that neglecting the radial expansion/contraction, the convective element moves as a rigid body, therefore any points of  its surface co-move with the stagnation points, i.e. $\theta=0, \phi=0$. Finally, the condition of quasi hydrostatic equilibrium applies, i.e. (${P \over \rho} + \Phi_g)\cong 0$. Eventually, the square of the velocity is (remember the definition of Eq.(\ref{Eq015}) as acceleration of the fluid as seen from $S_1$ onto the convective element surface)
\begin{eqnarray}\label{Eq049}
	{v^2} &=&  - A{\xi _e} \nonumber\\
				&=& \frac{{\nabla  - {\nabla _e} + \frac{\varphi }{\delta }{\nabla _\mu }}}{{\frac{{3{h_P}}}{{2\delta \Delta z}} + \left( {{\nabla _e} + 2\nabla  - \frac{\varphi }{{2\delta }}{\nabla _\mu }} \right)}}{\xi _e}g,
\end{eqnarray}
where we have used Eq.(\ref{Eq043}).  From the previous equation we obtain:
\begin{equation}\label{Eq050}
\frac{{3{h_P}}}{{2\delta v\Delta t}}\frac{{{v^2}}}{{{\xi _e}}} + \left( {{\nabla _e} + 2\nabla  - \frac{\varphi }{{2\delta }}{\nabla _\mu }} \right)\frac{{{v^2}}}{{{\xi _e}}} = \left( {\nabla  - {\nabla _e} - \frac{\varphi }{\delta }{\nabla _\mu }} \right)g,
\end{equation}
which, once the limit $\epsilon = v/\dot{\xi}_e \longrightarrow 0$, see  Eq.(\ref{Eq011}) and corollary 2, i.e. at $t \to \infty$ behaves as
\begin{equation}\label{Eq051}
\frac{{{v^2}}}{{{\xi _e}}} = \frac{{\nabla  - {\nabla _e} - \frac{\varphi }{\delta }{\nabla _\mu }}}{{{\nabla _e} + 2\nabla  - \frac{\varphi }{{2\delta }}{\nabla _\mu }}}g.
\end{equation}
This is a remarkable result that at a first sight may look surprising: the dominant term of the acceleration (in the denominator) $\frac{{3{h_P}}}{{2\delta v\Delta t}} > {\nabla _e} + 2\nabla  - \frac{\varphi }{{2\delta }}{\nabla _\mu }$ does not affect the velocity at the regime in which we are going to integrate our solution of the Navier-Stokes equation, see Eq.(\ref{Eq010}).

This counter-intuitive result is mathematically consistent only asymptotically in time and valid only within the approximation adopted in Eq.(\ref{Eq011}). It will be numerically checked in the next section (see Fig.\ref{FigQ}) where we will see that the ratio of the two terms of the LHS  of Eq.(\ref{Eq050}) shows a maximum not yet investigated in the literature. It  means that the mechanical evolution of a convective element (for instance its expansion) is dominated more by the local gradients of temperature over the pressure and less by its location inside the star ${\bm{x_e}}$. What we may learn from the above analysis is that the transfer of energy by convection is governed more by the expansion of the convective cells than by  their upward/downward motions. It follows from this, that the properties of convection  are mainly  driven by local rather than large scale physics. This lends support to the hypotheses  already implicit in the MLT.  However, we remark that the above locality does not contradict the spatial changing of  temperature gradients ($\nabla$, ${\nabla _e}$ etc.) and  gravitational force across the star.

Finally, recalling  that $\Delta T = \frac{T}{{{h_P}}}\left( {\nabla  - {\nabla _e}} \right)v\Delta t$ \citep[e.g. Eq. (6.19)][]{2013sse..book.....K}, we can define the convective flux along the radial direction as \footnote{This is derived from  Eq.(\ref{Eq049}) using this formalism and recalling  the fundamental theorem of calculus for a Lipschitzian function, according to which  $\Delta {{\bm{x}}_{O'}}$ differs from ${\bm{v}}\Delta t$ by the  quadratic terms $\textrm{O}{\left( {\Delta t} \right)^2}$, i.e. beyond the  approximation made in the Schwarzschild-Ledoux criteron or  Eq.(\ref{Eq044}).}
\begin{equation}\label{Eq052}
{\varphi _{{\rm{cnv}}}} = {1 \over 2} {c_P}{\rho }T\left( {\nabla  - {\nabla _e}} \right)\frac{{{\bar v ^2}\Delta t}}{{{h_P}}},
\end{equation}
where the factor 1/2 comes from the fact that at each level approximately one-half of the matter is rising and one-half is descending \citep{2004cgps.book.....W} to secure mass conservation locally.
Finally, recalling that the fluid acceleration in $S_0$ is seen as a negative quantity,  we can also write the flux as ${\varphi _{\rm{cnv}}} \propto   {c_P}{\rho }T\left( {{\nabla _e} - \nabla } \right) {\sqrt {A{\xi _e}} } \frac{{\Delta {{\bm{x}}_{O'}}}}{{{2 h_P}}}$, and we see that this equation is equivalent to the standard formalism, e.g. Eq.(7.7) of  \citet{1994sse..book.....K}, if we use the velocity derived from their heuristic considerations because the term ${\left( {A {\xi _e}} \right)^{1/2}}$ takes the dimension of a distance $\left[ D \right]$ over a velocity $\frac{{\left[ D \right]}}{{\left[ T \right]}}$; indeed $\left[ {{{\left( {A {\xi _e}} \right)}^{1/2}}} \right] = {\left( {\frac{{\left[ D \right]}}{{{{\left[ T \right]}^2}}}\left[ D \right]} \right)^{1/2}} = \frac{{\left[ D \right]}}{{\left[ T \right]}}$ as required by the dimensional analysis.

\subsection{The basic equations of stellar convection without the mixing-length parameter}\label{EqSc}
The ultimate result we are seeking is a self-consistent solution of the equations for the convective transport of energy  inside a convective layer, without  making use of  adjustable parameters such as the mixing-length ${\Lambda _m}$. For simplicity we present here the  case of  a chemically homogeneous medium ${\nabla _\mu } = 0$, since in any case the major role of the convection is indeed to homogenize gradients of chemical composition. The general case ${\nabla _\mu }\left( {\bm{x}} \right) \ne 0$ comes then trivially with a suitable change of variables as will be more evident a posteriori. The key result, and ultimately one of the goals of this paper is to prove that if we call the body of variables (temperature, pressure, density, etc..) defining the physical state of stellar interiors at a given position ${\bm{x}}$ the ``stellar-system'', the following holds:

\textbf{Theorem of the uniqueness  of the stellar convection.} \textit{The radiative  ${\nabla _{{\rm{rad}}}}$, the adiabatic ${\nabla _{{\rm{ad}}}}$, the local gradient of the star $\nabla $, and the convective element gradient ${\nabla _e}$  are in a one-to-one correspondence (a bijection) with the stellar system in which they are embedded.}

\textbf{Proof}: To prove the assertion of this theorem we need to solve the equation of stellar convection without any free parameter (e.g. the mixing-length  ${\Lambda _m}$) thus unequivocally assigning  to each location inside a star its own characteristic convection. In other words,  we are going to describe the stellar convection not as a one-parameter family of solutions (i.e. the mixing-length parameter ${\Lambda _m}$ to be fixed by external constraints) but with a unique solution of the system of equations governing stellar convection.
We start by extending  the present formalism to include a few fundamental theoretical tools.
A convective cell of mass ${m_e}$, volume ${\upsilon _e}$ and radius ${\xi _e}$,  once it has acquired a positive excess of temperature $|\Delta T| = T \left( {\nabla  - {\nabla _e}} \right)\frac{{\Delta {{\bm{x}}_{O'}}}}{{{h_P}}}$ with respect to its surroundings, radiates energy into the stellar medium with a flux ${\varphi _{\rm{rad}}} = \frac{{4ac{T^3}}}{{3\kappa \rho }}\left| {{\nabla _{{\bm{\hat n}}}}T} \right|$, where  $a =7.5657 \times {10^{ - 16}}{\rm{J}}{{\rm{m}}^{{\rm{ - 3}}}}{{\rm{K}}^{{\rm{ - 4}}}}$ is the radiation-density constant, $\kappa $ the mean absorption coefficient, or opacity, and $c$ the speed of light. The radiative loss per unit of time $\frac{{d{Q_{{\rm{loss}}}}}}{{dt}}$ from the convective element from this radiative flux and  its adiabatic expansion  causes a temperature decrease, simply because from Eq.(\ref{Eq046})    $d{Q_{{\rm{loss}}}} =  - {m_e}{c_P}dT \Rightarrow {\dot Q_{{\rm{loss}}}} = \rho_e {\upsilon _e}{c_P}\left\langle{\nabla _{\bm{x}}}T,{\bm{\dot x}}\right\rangle $. We then relate the radiative loss $\frac{{d{Q_{{\rm{loss}}}}}}{{dt}} = \frac{{8ac{T^3}}}{{3\kappa \rho }}T\left( {\nabla  - {\nabla _e}} \right)\frac{|{\Delta {\bm{x}}|}}{{{h_P}}}\frac{S}{{2{\xi _e}}}$ to the temperature gradient  $||{\nabla _{\bm{x}}}T ||=  - \frac{1}{{\rho V{c_P}v}}\frac{{d{Q_{{\rm{loss}}}}}}{{dt}}$  using the formalism of Section \ref{EqcnvQ} by recalling that   $\frac{\Sigma}{{2V{\xi _e}}} = \frac{3}{{2\xi _e^2}}$ (with $V$ and $\Sigma$ volume and surface of the convective element) to obtain the relation
\begin{equation}\label{Eq053}
\frac{{{\nabla _e} - {\nabla _{{\rm{ad}}}}}}{{\nabla  - {\nabla _e}}} = \frac{{4ac{T^3}}}{{\kappa {\rho ^2}{c_P}}}\frac{\Delta t }{{\xi _e^2}},
\end{equation}
which represents another equation to solve together with those we have developed.

Furthermore, in addition to the convective flux one should consider the flux carried by radiation and conduction. The radiative flux is ubiquitous and no other comments are necessary. Suffice to recall here that it depends on the temperature gradient existing in the stellar medium and the so-called Rosseland mean opacity.  Conduction has an important role in the degenerate cores of red giants and advanced stages of intermediate-mass and massive stars, and dominates  in the isothermal cores of white dwarfs  and neutron stars. The conductive flux can be expressed by the same relation for the radiative flux provided the opacity is suitably redefined.
In the following we will limit ourselves to the case of normal (main sequence) stars and therefore leave conduction aside. However to consider the possibility of including the conductive flux, we indicate with ${\varphi _{{\rm{rad|cnd}}}}$ either the radiative flux alone or the radiative and conductive fluxes lumped together with the mean opacity $\kappa$ suitably redefined  \citep[see][for all details]{2013sse..book.....K}.
Therefore, in our simplified situation, the total flux is the sum of the radiative and convective terms ${\varphi _{{\rm{rad}}}}+{\varphi _{{\rm{cnv}}}}$.

We define  now the gradient ${\nabla _{{\rm{rad}}}}$ that would be necessary to transport the \textit{total} flux by radiation alone as:
\begin{eqnarray}\label{Eq054}
	{\varphi _{{\rm{rad}}}} + {\varphi _{{\rm{cnv}}}} &=& \frac{{4acG}}{3}\frac{{{T^4}m}}{{\kappa p{{\left\| {{{\bm{x}}_{O'}}} \right\|}^2}}}{\nabla _{{\rm{rad}}}} \nonumber\\
	&=& \frac{{4ac}}{3}\frac{{{T^4}}}{{\kappa {h_P}\rho }}{\nabla _{{\rm{rad}}}}.
\end{eqnarray}
Denoting with $\nabla $ the ambient gradient in presence of radiation and convection  the amount of energy carried by the sole radiation (or radiation + conduction) is
\begin{eqnarray}\label{Eq055}
	{\varphi _{{\rm{rad|cnd}}}} &=& \frac{{4acG}}{3}\frac{{{T^4}m}}{{\kappa P{{\left\| {{{\bm{x}}_{O'}}} \right\|}^2}}}\nabla \nonumber\\
	&=& \frac{{4ac}}{3}\frac{{{T^4}}}{{\kappa {h_P}\rho }}\nabla.
\end{eqnarray}
We recollect here the system of six equations Eqs. (\ref{Eq055}), (\ref{Eq054}), (\ref{Eq053}), (\ref{Eq052}), (\ref{Eq051}), and (\ref{Eq015}) (considered with the non-dimensional numerical solution of $\chi$, Eq.(\ref{Eq022}), and with its dimensional form Eq.(\ref{Eq019})) of the six unknowns $\left\{ {{\varphi _{{\rm{rad|cnd}}}},{\varphi _{{\rm{cnv}}}},v,{\nabla _e},\nabla ,{\xi _e}} \right\}$ that we solve as a function of position inside the star ${\bm{x}}$ and for ${t \to \infty }$, once the quantities $\left\{ {P,T,\rho ,l,m,{c_P},{\nabla _{{\rm{ad}}}},{\nabla _{{\rm{rad}}}},g} \right\}$ or $\left\{ {P,T,\rho ,l,m,{c_P},{\nabla _{{\rm{ad}}}},{\nabla _{{\rm{rad}}}},{\nabla _\mu },g} \right\}$ are locally known as function of ${\bm{x}}$. In the case of a chemically homogeneous  layer unstable to convection, the system of equations for $t \to \infty $ is\footnote{Since we are interested only in the asymptotic behaviour of the system we can insert Eq.(\ref{Eq044}) already acconting for its asymptotic behaviour, see the remarks  on  Eq.(\ref{Eq044}) and Eq.(\ref{Eq051}) for chemically homogeneous layers. However, when performing the  numerical integration presented in Section \ref{NumEq}, all terms will  be included.}:

\begin{equation}\label{Eq056}
\left\{ \begin{array}{rcl}
{\varphi _{\rm{rad|cnd}}} &=& \frac{{4acG}}{3}\frac{{{T^4}m}}{{\kappa P{\left\| {{{\bm{x}}_{O'}}} \right\|^2}}}\nabla   \\
{\varphi _{\rm{rad|cnd}}} + {\varphi _{\rm {cnv}}} &=& \frac{{4acG}}{3}\frac{{{T^4}m}}{{\kappa P{\left\| {{{\bm{x}}_{O'}}} \right\|^2}}}{\nabla _{\rm{rad}}} \\
\frac{{{{\bar v}^2}}}{\bar\xi_e} &=& \frac{{\left( {\nabla  - {\nabla _e}} \right)}}{{\left( {{\nabla _e} + 2\nabla } \right)}}g\\
{\varphi _{\rm{cnv}}} &=& {1\over 2}  \rho {c_p}T\left( {\nabla  - {\nabla _e}} \right)\frac{{{{\bar v}^2}{t_0}\tau }}{{{h_p}}}\\
\frac{{{\nabla _be} - {\nabla _{\rm{ad}}}}}{{\nabla  - {\nabla _e}}} &=& \frac{{4ac{T^3}}}{{\kappa {\rho ^2}{c_p}}}\frac{{{t_0}\tau }}{{{\bar\xi_e ^2}}}\\
\bar\xi_e  &=& {\left( {\frac{{{t_0}}}{2}} \right)^2}\frac{{\nabla  - {\nabla _b}}}{{{\nabla _b} + 2\nabla }}g\bar\chi \left( \tau  \right)
\end{array} \right.
\end{equation}
where the last equation is the convective element equation studied in Section \ref{EqCnv}, time averaged (see Eq.(\ref{Eq015}) and Eq.(\ref{Eq019}) and Section \ref{convflux}).
To prove the theorem we need to show that the asymptotic behaviour of this system of equations is time independent, i.e. we do not need to introduce any temporal time-scale (or any arbitrary  spatial scale $l_m$ as required by the MLT), i.e. the solution of the system is a unique manifold. This will induce an asymptotic behaviour in the numerical solution of the system, which  will be presented  in the Section \ref{NumEq}.

\begin{figure}
\includegraphics[width=\columnwidth]{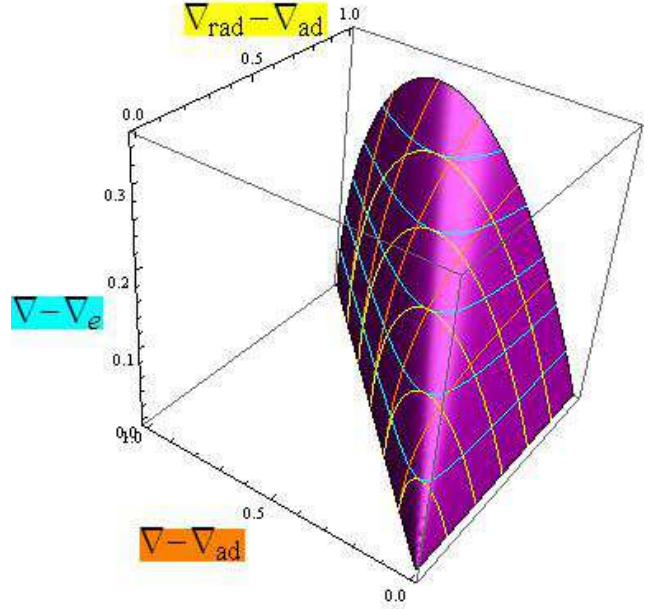}
\caption{The manifold represent the surface of all the possible solution for the convection equation. Surfaces of constant $W$, $\eta$ and $Y$ are plotted but labelled with their corresponding definitions: $W$ in yellow, $\eta$ in orange and $Y$ in blue. The purple manifold solution is computed for all the numerical values of interest for stars on the left-side of the Hayashi-line in the HR-diagram.}%
\label{Fig3}
\end{figure}

The solution of this set of algebraic equations leads to a manifold that determines the gradients for which we are seeking, i.e. $\nabla_e $ and $\nabla$. We proceed to this solution in the next section, here we prove that the relation between these two gradient is unique as follows. We substitute the first equation into the second one to reduce the number of equations from six to five. We then substitute its result into the third equation thus obtaining a set of four equations
\begin{equation}\label{Eq057}
\left\{ \begin{array}{rcl}
	{\varphi _{{\rm{cnv}}}} &=& \frac{{4ac{T^4}}}{{3\kappa {h_P}\rho }}\left( {{\nabla _{{\rm{rad}}}} - \nabla } \right)\\
	{\varphi _{{\rm{cnv}}}}\left( {{\nabla _e} + 2\nabla } \right) &=& \frac{{\rho {c_P}T\tau g}}{{{h_P}}}{\xi _e}{\left( {\nabla  - {\nabla _e}} \right)^2}\\
	\frac{{{\nabla _e} - {\nabla _{{\rm{ad}}}}}}{{\nabla  - {\nabla _e}}} &=& \frac{{4ac{T^3}}}{{\kappa {\rho ^2}{c_P}}}\frac{\tau }{{\bar\xi _e^2}}\\
	{\bar\xi _e} &=& \frac{g}{4}\frac{{\nabla  - {\nabla _e}}}{{{\nabla _e} + 2\nabla }} \bar \chi.
	\end{array} \right.
\end{equation}
Inserting now   the first equation into the second one and defining  two  auxiliary quantities  depending only on the local properties of the star
\begin{equation}\label{Eq058}
\begin{array}{*{20}{c}}
	{k \equiv \frac{{ac{T^3}}}{{\kappa {\rho ^2}{c_P}}}}&\,\,\,\,\,\,\,{\rm and } \,\,\,\,\,\,\,{{g_4} \equiv \frac{g}{4}}
	\end{array},
\end{equation}
we get:
\begin{equation}\label{Eq059}
\left\{ \begin{array}{rcl}
	\frac{{{{\left( {\nabla  - {\nabla _e}} \right)}^2}}}{{\left( {{\nabla _e} + 2\nabla } \right)\left( {{\nabla _{{\rm{rad}}}} - \nabla } \right)}}{\bar\xi _e} &=& \frac{k}{{3\tau {g_4}}}\\
	\frac{{{\nabla _e} - {\nabla _{{\rm{ad}}}}}}{{\nabla  - {\nabla _e}}}\bar\xi _e^2 &=& \tau k\\
	\frac{{{\nabla _e} + 2\nabla }}{{\nabla  - {\nabla _e}}}{\bar\xi _e} &=& {g_4} \bar\chi.
	\end{array} \right.
\end{equation}
Furthermore,  taking the ratio of the first to the third equation and the ratio of second equation to the square of the third one, after some algebraic manipulations  we get:
\begin{equation}\label{Eq060}
\left\{ \begin{array}{rcl}
	\frac{{{{\left( {\nabla  - {\nabla _e}} \right)}^3}}}{{{{\left( {{\nabla _e} + 2\nabla } \right)}^2}\left( {{\nabla _{{\rm{rad}}}} - \nabla } \right)}} &=& \frac{1}{{3\tau }}\frac{k}{{g_4^2\chi }}\\
	\frac{{\left( {{\nabla _e} - {\nabla _{{\rm{ad}}}}} \right)\left( {\nabla  - {\nabla _e}} \right)}}{{{{\left( {{\nabla _e} + 2\nabla } \right)}^2}}} &=& \frac{\tau }{\chi }\frac{k}{{g_4^2 \bar\chi }}.
	\end{array} \right.
\end{equation}
For each layer inside the convectively unstable region, we define a few auxiliary variables:
\begin{equation}\label{Eq061}
W \equiv {\nabla _{{\rm{rad}}}} - {\nabla _{{\rm{ad}}}} > 0,
\end{equation}
and
\begin{equation}\label{Eq062}
\eta  \equiv \nabla  - {\nabla _{{\rm{ad}}}},
\end{equation}
\begin{equation}\label{Eq063}
Y \equiv \nabla  - {\nabla _e}.
\end{equation}
Using these expressions we can write
\begin{eqnarray}
  {\nabla _{{\rm{rad}}}} - \nabla     &=& W - \eta             \nonumber    \\
  {\nabla _e} - {\nabla _{{\rm{ad}}}} &=& \eta  - Y            \nonumber    \\
  {\nabla _e} + 2\nabla  &=& - Y + 3\left( {\eta  + {\nabla _{{\rm{ad}}}}} \right) \label{Eq064}.
\end{eqnarray}
Finally  Eq.(\ref{Eq060})  yields the most important  relation and result of our study. Merging the two equations we get:
\begin{equation}\label{Eq065}
\frac{{{Y^2}}}{{\left( {W - \eta } \right)\left( {\eta  - Y} \right)}} = \frac{1}{3}\frac{\bar \chi }{{{\tau ^2}}}.
\end{equation}
which we need to solve for $\tau  \to \infty$. But recalling  Eq.(\ref{Eq031}), the  asymptotic temporal dependence of this relation (RHS $\rightarrow \rm{const.}$ for $\tau \rightarrow \infty $) establishes that   convection inside stars \textit{does not depend on time evolution and/or  any spatial scale parameter, to first order} (i.e. it is independent from any the mixing-length/mixing-time):
\begin{equation}\label{Eq066}
\frac{{{Y^2}}}{{\left( {W - \eta } \right)\left( {\eta  - Y} \right)}}{\rm{ = const}}.
\end{equation}
This equation in the space of $W$, $\eta$ and $Y$ describes a surface containing the manifold of all possible solutions. This manifold is graphically represented  in Fig. \ref{Fig3}, where  $W$, $\eta $ and $Y$ are replaced by their definitions: Eq.(\ref{Eq061}), (\ref{Eq062}) and (\ref{Eq063}) and the RHS constant have been evaluated at some arbitrary layer inside the convective region. It is worth recalling here that of the four temperature gradients that are involved, i.e. $\nabla_{\rm{rad}}$, $\nabla_{\rm{ad}}$, $\nabla_e$, $\nabla$, the adiabatic gradient $\nabla_{\rm{ad}}$ is always known given the thermodynamical state of the medium, and the radiative gradient $\nabla_{\rm{rad}}$ is known once the total flux is specified (this is the typical case of convection in the outer layers, where the MLT and/or the present theory are best suited). We are left with the unknown gradients $\nabla_e$ and $\nabla$ asymptotically related by an unique relation, i.e. all the unknowns of the system are in a one-to-one correspondence without any free parameter (Q.E.D.). 

It goes without saying that a different free-parameter manifold can be worked out to prove the theorem, and the constant at RHS of Eq. (\ref{Eq066}), $\nabla_{\rm{rad}}$ and $\nabla_{\rm{ad}}$ depend on the position inside the star, so that each layer has its own values for $\nabla_e$ and $\nabla$. The study of the solution for all the unknowns of our original system is presented in the next section.

Finally,  in the case of a chemically non homogeneous medium, ${\nabla _\mu } \ne 0$, after  changing the definition of $Y$ in Eq. (\ref{Eq063}) to
\begin{equation}\label{Eq067}
Y \equiv \nabla  - {\nabla _e} + \frac{\varphi }{\delta }{\nabla _\mu },
\end{equation}
we obtain a solution manifold in the form of Eq.(\ref{Eq066}) (but with a different constant). Thus, an analogous theorem holds in the case of a chemically non-homogeneous convectively-unstable layer. This represents the mathematical proof of the recent finding in numerical investigations by \citet{2013ApJ...767...78T}.

\subsection{Numerical solution: comparing the new theory with the classical MLT}\label{NumEq}

The previous theorem immediately suggests a time-independent behaviour for the functions that are solutions of our system of equations for stellar convection. Thus, we expect a numerical integration of the system of equations
 \begin{equation}\label{Eq068}
\left\{ \begin{array}{rcl}
	{\varphi _{{\rm{rad|cnd}}}} &=& \frac{{4ac}}{3}\frac{{{T^4}}}{{\kappa {h_P}\rho }}\nabla \\
	{\varphi _{{\rm{rad|cnd}}}} + {\varphi _{{\rm{cnv}}}} &=& \frac{{4ac}}{3}\frac{{{T^4}}}{{\kappa {h_P}\rho }}{\nabla _{{\rm{rad}}}}\\
	\frac{{{\bar{v}^2}}}{{{4\xi _e}}} &=& \frac{{\nabla  - {\nabla _e}}}{{{3h_P \over 2\delta \bar{v} \tau}+{\nabla _e} + 2\nabla }}g_{4}\\
	{\varphi _{{\rm{cnv}}}} &=&{1 \over 2} \rho {c_P}T\left( {\nabla  - {\nabla _e}} \right)\frac{{{\bar{v}^2}\tau }}{{{h_P}}}\\
	\frac{{{\nabla _e} - {\nabla _{{\rm{ad}}}}}}{{\nabla  - {\nabla _e}}} &=& \frac{{4ac{T^3}}}{{\kappa {\rho ^2}{c_P}}}\frac{{\tau }}{{\bar\xi _e^2}}\\

	{\bar\xi _e} &=& \frac{{\nabla  - {\nabla _e}}}{{ {3h_P \over 2\delta \bar{v} \tau} + {\nabla _e} + 2\nabla }}
g_4{\bar\chi} \left( \tau  \right),
	\end{array} \right.
\end{equation}
to present an asymptotic behaviour in time for at least some of the function solutions. In Eq. (\ref{Eq068}) the last equation is obtained from Eq.(\ref{Eq051}) with $\nabla_\mu=0$ and the help of Eq. (\ref{Eq019}). Note that in the third and sixth equation of the system, the term ($3 h_p / 2\delta \bar{v} \tau$) is retained.
Although an algebraic solution of this system (containing 6 equations and 6 unknowns as function of the time) can be worked out with an algebraic manipulator, it is exceptionally long and does not add a better comprehension of the physics we are investigating. For this reason, we present here a numerical investigation that enlightens the features of the system we presented and deferring to a future study the investigation of the complete ADE system (Pasetto et al. 2014b in preparation).

At the same time, this will prove that we obtain correct numerical values for  the gradients studied  and  that we
can  directly compare our results  with the standard ones of the literature based on complete stellar models calculated  with any of the sophisticated codes of
stellar structure in the literature, e.g. the classical G\"{o}ttingen code developed by \citet{Hofmeister1964}, the many versions of this developed over the years by the Padova group, e.g. \citet{ChiosiSumma1970} with semi-convection, \citet{1981A&A...102...25B} with ballistic convective overshoot from the core, \citet{Alongi1991} with envelope overshoot, \citet{Deng1996a,Deng1996b,Salasnich1999} with turbulent diffusion,
finally the many revision and improvements described in \citet{1994A&AS..106..275B, 1995A&A...301..381B, 2008A&A...484..815B, 2001AJ....121.1013B,2003AJ....125..770B, 2009A&A...508..355B}, and the Garching version developed by \citet[][GARSTEC]{Weiss2008}.

Using the library of  complete stellar models of  \citet{2009A&A...508..355B} for different values of the
stellar mass and chemical composition, calculated with the standard MLT ($\Lambda_m=1.64$),  we select a typical model for the Sun  on which we can test the new theory of convection in a very simple way while we leave an extended numerical investigation of a different stellar models to a future study \citep[][in preparation]{Pasetto14}.
Indeed the Sun is the best place to test the new theory of convection because for it we have the most complete information \citep[see, e.g.][and references therein]{2012ApJ...755L..12B} .
The solar model provides the mass, luminosity, pressure, density, temperature, opacity, chemical composition and many other physical quantities  throughout the Sun and  we have precise data on the total luminosity, effective temperature and radius in addition to surface abundances and a rather precise estimate of the age.

In particular the Sun's model provides us with $\nabla_{\rm{rad}}$ and $\nabla_{\rm{ad}}$ the two gradients that are needed to start the analysis and that do not depend on the convection theory in use. It is worth recalling here that $\nabla_{\rm{ad}}$ in presence of ionization, as it occurs in the external layers of a star, is a complicated function of EoS, temperature, density, degree of ionization, etc. that cannot be approximated by  simple analytical expressions. It becomes a simple function of the EoS only when ionization is complete \citep{2004cgps.book.....W}. The model of the Sun we are using includes ionization and takes it into account when calculating $\nabla_{\rm{ad}}$.

Using this model we calculate  $\nabla$ and $\nabla_e$, velocities etc. both according to the new formalism for convection and also to the standard MLT in the \citet{2009A&A...508..355B} model adopting for MLT the current estimate for $\Lambda_m$ in the Sun, i.e. $\Lambda_m=1.64$.

\begin{figure}
\includegraphics[width=\columnwidth]{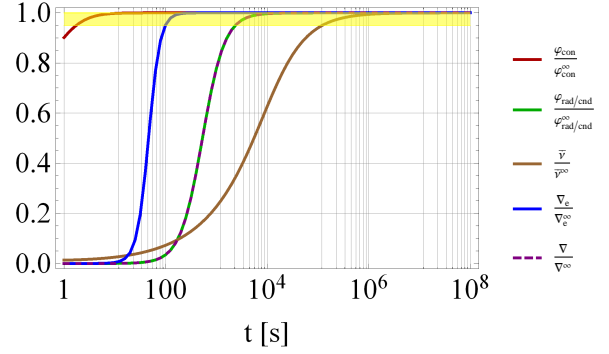}
\caption{Temporal dependence of the mean velocity of a convective element in the Sun at the layer
$z= \frac{98}{100}R_\odot$ . Remarkably, the velocity reaches the asymptotic value within a rather short time scale (of the order of a month or so). The yellow bar shows the 5\% region of the asymptotic value of the various quantities on display.
\label{Fig4_bis}}
\end{figure}

Given  the gradients $\nabla_{\rm{ad}}$ and $\nabla_{\rm{rad}}$, together with the function $\bar\chi(\tau)$, in each layer in the external convective regions of the Sun, we solve the system (\ref{Eq068}) as a function of $\tau$ (i.e. time) and position to derive the gradients $\nabla$ and $\nabla_e$, the mean velocity $\bar v$, and the convective flux $\varphi_{\rm{cnv}}$. Since the term ($3 h_p / 2\delta \bar{v} \tau$) is retained, the system will relax to that of Eqs. (\ref{Eq056}) only after a certain time interval has elapsed. The best way of evaluating how long the time interval necessary to reach the asymptotic behaviour is, is to plot the time dependence of the velocity, the temperature gradients $\nabla$ and $\nabla_e$, and the convective fluxes $\varphi_{\rm{cnv}}$, $\varphi_{\rm{rad/cnd}}$ (Fig. \ref{Fig4_bis}). In particular in Fig. \ref{Fig4_bis} the yellow bar indicates when the asymptotic values are reached within 5\% of their limit value. These curves plotted refer to the layer $z= \frac{98}{100}R_\odot$, i.e. a shell representative of the  external  convective region of the Sun.  All the quantities of interest reach the asymptotic value on a time scale of one month or even shorter.

Note that with our theory the path travelled by the bubble is of course not defined, nor has it any physical meaning, being a theory developed in a system of reference $S_1$  where the bubble is at rest by definition. 

Finally, it might be of interest to estimate the typical lifetime $t_*$ of a bubble by taking the ratio of the natural scale length $h_p$ to typical expansion velocity of a convective cell  ${t_*} \equiv \frac{{{h_P}}}{{{\dot \xi _e}}}$ that at $z = {\rm{98 \over 100}}{R_ \odot }$ is ${t_*} \sim 3.2$ hours.

\begin{figure}
\includegraphics[width=\columnwidth]{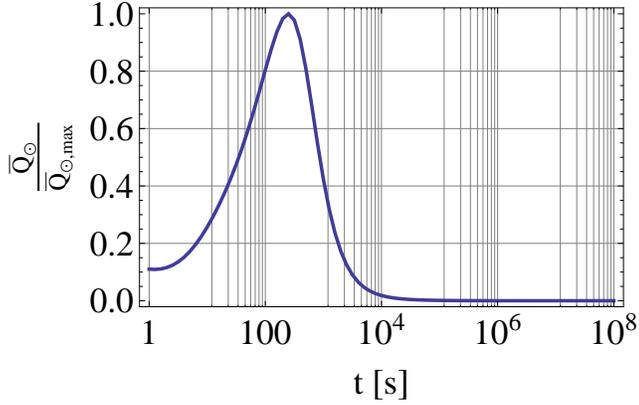}
\caption{Q parameter for the model of the Sun in an arbitrary but fixed point in the Sun at the layer
$z= \frac{98}{100}R_\odot$. The plot was normalized to the value of the maximum.}
\label{FigQ}
\end{figure}

Now we verify the assumption leading to our Eq.(\ref{Eq050}) with a direct numerical integration  of this equation as a function of time, using the input physics of Sun model. If we define the ratio between the first and second term in the LHS of Eq.(\ref{Eq050}) computed with the average velocity of the convective elements:
\begin{equation}
{\bar Q_ \odot } \equiv \frac{{\frac{{3{h_P}}}{{\bar v\Delta t2\delta }}}}{{{\nabla _e} + 2\nabla }},
\end{equation}
we have seen that the asymptotic behaviour expected as $t \to \infty$ requires this ratio to converge to zero. The same trend is expected for the average behaviour of the convective elements. What is not expected a priori from a simple asymptotic expansion is the maximum that we see in Fig.\ref{FigQ}.
This maximum is the result of the opposite time dependence of the numerator and denominator: while the denominator progressively increases toward its asymptotic value as shown in Fig. \ref{Fig4_bis}, the numerator monotonically decreases with time.

In column (3) Table \ref{table:1} we list the results we have obtained from the system of Eq.(\ref{Eq068}) limited to the layer we have selected ($z= \frac{98}{100}R_\odot$). Any other convectively-unstable layer would  have shown similar results. A numerical investigation of the consequences of the present theory and a complete upgrade of the stellar models in \cite{2008A&A...484..815B,2009A&A...508..355B} is deferred to the forthcoming paper \cite{Pasetto14}.

Now we compare our results  with those obtained from the standard MLT  of stellar convection represented by the system of equations
\begin{equation}\label{Eq069}
\left\{ \begin{array}{rcl}
{\varphi _{{\rm{rad|cnd}}}} &=& \frac{{4ac}}{3}\frac{{{T^4}}}{{\kappa {h_P}\rho
}}\nabla \\
{\varphi _{{\rm{rad|cnd}}}} + {\varphi _{{\rm{cnv}}}} &=&
\frac{{4ac}}{3}\frac{{{T^4}}}{{\kappa {h_P}\rho }}{\nabla _{{\rm{rad}}}}\\
{{\bar v}^2} &=& g\delta \left( {\nabla  - {\nabla _e}} \right)\frac{{l
_m^2}}{{8{h_P}}}\\
{\varphi _{{\rm{cnv}}}} &=& \rho {c_P}T\sqrt {g\delta } \frac{{l _m^2}}{{4\sqrt
2 }}h_P^{ - 3/2}{\left( {\nabla  - {\nabla _e}} \right)^{3/2}}\\
\frac{{{\nabla _e} - {\nabla _{{\rm{ad}}}}}}{{\nabla  - {\nabla _e}}} &=&
\frac{{6ac{T^3}}}{{\kappa {\rho ^2}{c_P}{l_m}{\bar v}}},
\end{array} \right.
\end{equation}
in which $l_m$  contains the mixing-length parameter $\Lambda_m$. The  derivation and solution  of this system of equations can be found in  any classical textbook of stellar structure \citep[e.g.][]{ 2013sse..book.....K, 2004cgps.book.....W}. We limit ourselves to note that in this classical system we have five equations instead of six, see  Eq.(\ref{Eq056}). If we adopt the same model of the Sun we have used before to solve the system
Eq.(\ref{Eq056})  with the extra value of $\Lambda_m$ tuned on the Sun,  we obtain the results
presented in  column (4) of Table \ref{table:1}. The results are practically coincident with those from this new theory.

\begin{figure}
\includegraphics[width=\columnwidth]{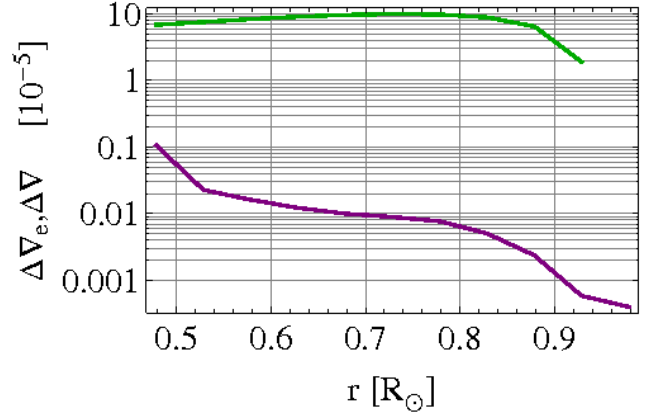}
\caption{Normalized difference function evaluated for ${\nabla _{{\rm{e}}}}$ ($\Delta{\nabla _{{\rm{e}}}}$, purple line) and ${\nabla _{{\rm{}}}}$ (${\Delta\nabla _{{\rm{}}}}$, green line). MLT values were computed assuming ${\Lambda _m} = 1.64$}%
\label{Fig10}
\end{figure}

\begin{figure}
\includegraphics[width=\columnwidth]{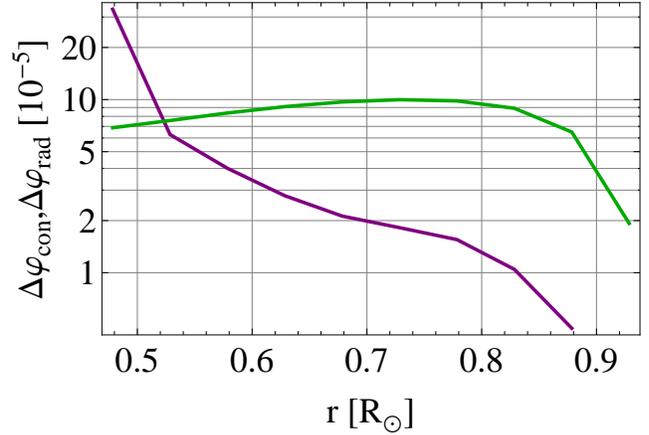}
\caption{Normalized difference function evaluated for convective ($\Delta{\varphi _{{\rm{cnv}}}}$, purple line) and radiative flux ($\Delta{\varphi _{{\rm{rad}}}}$, green line). Same ${\Lambda _m}$ has been adopted as Fig.\ref{Fig10}}%
\label{Fig11}
\end{figure}

The comparison between our theory and the standard MLT predictions can then be extended over the entire convective region   inside the Sun. We define a normalized difference function as $\Delta \Xi \left( {\bm{x}} \right) \equiv \left| {\frac{{{\Xi _{{\rm{MLT}}}}\left( {\bm{x}} \right) - {\Xi _{{\rm{new}}}}\left( {\bm{x}} \right)}}{{{\Xi _{{\rm{new}}}}\left( {\bm{x}} \right)}}} \right|$ where $\Xi  = {\nabla _{{\rm{}}}},{\nabla _e}$, etc., i.e. for every function of interest we compute the difference of its values obtained with the standard MLT, ${\Xi _{{\rm{MST}}}}$, and our new approach ${\Xi _{{\rm{new}}}}$. The results are plotted in Fig.\ref{Fig10} for $\Xi  = {\nabla _{{\rm{}}}}$ and $\Xi  = {\nabla _{{\rm{e}}}}$ and in Fig.\ref{Fig11} for $\Xi  = {\varphi _{{\rm{cnv}}}}$ and $\Xi  = {\varphi _{{\rm{rad}}}}$.

As it is evident from these figures,  the normalized differences between the two theoretical predictions are of the order of $O(10^{-5})$ over all the stellar radii of interest. This result holds independently from the stellar model adopted.

\begin{figure}
\includegraphics[width=\columnwidth]{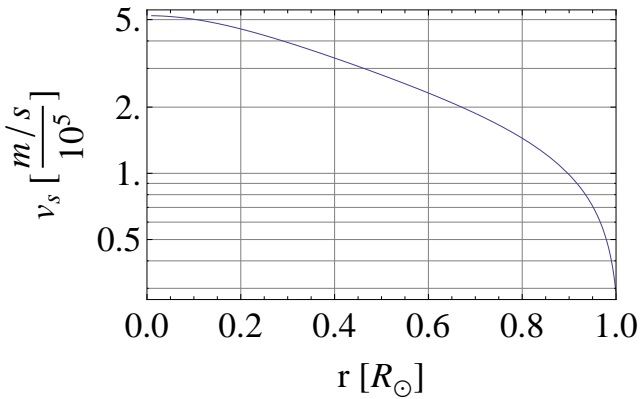}
\caption{Profile of the sound velocity in the adopted model of the Sun and limited to  the outer layers.}
\label{FigVs}
\end{figure}

Furthermore, whe show in Fig.\ref{FigVs} that the conditions that form the foundation of our theory (Eq.(\ref{Eq011})) are fully satisfied. The sound speed profile of our model, which is
part of our Eq.(\ref{Eq011}), is shown in Fig.\ref{FigVs}. The result is closely consistent with the models available from the literature \cite[e.g.][]{2004cgps.book.....W} or computed from \cite{2008A&A...484..815B}.

The different  velocities of the  convective elements predicted by the two theories may have some implications on the extension and efficiency of convective overshooting. The subject will be investigated in a forthcoming paper \citep[][in preparation]{Pasetto14}. Moreover, the meaning of the time-scales also differs. As far as the lifetime of the typical convective element is concerned at a chosen stellar radius, we evaluate this for the MLT as customary with ${t_*} = \frac{{{\Lambda _m}{h_P}}}{v} \sim 2.3$ days.

Finally, it is  worth remarking that from the mathematical point of view, the results do not depend on the particular stellar model we have examined.
Indeed, it would suffice to assign reasonable values to the
gradients  ${\nabla _{{\rm{rad}}}}$ and ${\nabla _{{\rm{ad}}}}$  and all other quantities to verify that \textit{both systems of equations, i.e. Eq.(\ref{Eq069}) or Eq.(\ref{Eq056}),  would lead to the same solutions even though the equations are different in form and physical meaning}.
This is confirmed by the entries of Table \ref{table:1}  and \textit{ by the difference  functions} shown in  Figs.\ref{Fig10} and \ref{Fig11}.

The results we have obtained come out in a simple and straightforward way. The physical foundations of the theory are simple
and free of ad hoc assumptions. Equally, this is the case for the mathematical developments that are carried out to reach the final result.

\begin{table}
\caption{Numerical results at the layer  ${{\bm{x}}_{O'}}= \frac{98}{100}{R_ \odot }$ inside the Sun's model. The differences between our theory and MLT at any other arbitrary but
fixed unstable convective layer are of the same order.}
\label{table:1}
\centering
\begin{tabular}{c c c c}
\hline\hline
                                    & Units                                                   & Eqs.(\ref{Eq068})       & Eqs.(\ref{Eq069})    \\
                                    &                                                         &                         & with ${\Lambda _m} = 1.64$\\
\hline
${\varphi _{{\rm{rad|cnd}}}}$       & ${\rm{erg}}\;{{\rm{s}}^{{\rm{ -1}}}}$     						  & 61.6629      & 61.7020 \\
${\varphi _{{\rm{cnv}}}}$           & ${\rm{erg}}\;{{\rm{s}}^{{\rm{ -1}}}}\times{10^7}$       & 6.42855      & 6.42854 \\
$h_P$                               & ${\rm{m}} \times {10^6}$                                & 2.13852      & 2.13852 \\
$\nabla_{\rm{rad}}$                 &                                                         & 295187.      & 295187. \\
    ${\nabla _e}$                   &                                                         & 0.28310      & 0.28310 \\
    $\nabla $                       &                                                         & 0.28315      & 0.28332 \\
    $\bar v$                             & ${\rm{m}}\;{{\rm{s}}^{ - 1}}$                      & 184.042      & 208.913 \\
\hline
\end{tabular}
\end{table}\label{Tab1}

\section{Conclusion and critical comments }\label{Conc}
It has taken almost one century to develop a theory for stellar convection and energy transport without the mixing-length parameter. In this first study we have presented a new simple theory of stellar convection that does not contain adjustable parameters such as the mixing-length. The whole solution (temperature gradients of the medium and convective elements, the distances travelled by typical elements, their velocities and lifetimes, the convective flux etc.) are all determined by the physical conditions inside the stars. We consider this to be a significant advance.

We have formulated the equations of fluid dynamics in the potential-flow approximation.
A  posteriori it is evident that  this is advantageous, simply because for a body rapidly expanding from rest it is a good approximation every time the inertia forces are larger than the viscous ones (at least on a
time-scale of the order of $\tau  \sim \textrm{O}\left( {\frac{{\Delta z}}{{\dot \xi }}} \right)$). This justifies this approximation and the description of the mechanics
of the convective elements that follows.

We summarise here the major features and the major achievements of our theory.
The approach is based on the addition of an equation for the motion of the convective
elements to the classical system of algebraic equations for the convective energy transport.
The motion of a convective element is described by the  vertical displacement of its barycenter and relative expansion (contraction) of its radius, and the inertia of the fluid mass displaced by the convective element is accounted for.
Consequently the acceleration imparted to the convective elements in addition to the buoyancy force takes into account effects that in the standard MLT are neglected, i.e. the inertial term of the fluid displaced by the movement and expansion (contraction) of the convective cell, and an extra term arising from the changing size of the convective element (the larger the convective element the stronger is the buoyancy effect and  the larger the acquired velocity and vice versa). This results in a new and more complicated term of the acceleration $\propto \frac{{\nabla  - {\nabla _e}}}{{\nabla  + 2\nabla }} \propto \nabla  - {\nabla _e}$, agreeing with the Schwarzschild criterion.

It is found that the best reference  frame to describe the system is the one comoving with the element. Our treatment of the fluid-dynamics governing the motion of   the convective elements allows us to remove any preliminary assumptions about the size and path of the convective elements and these now arise as natural outputs of our theory.

No external calibration of parameters is required: the solution of the equations governing stellar convection is unique, in the sense that it is fully determined by  physical properties of the medium. This is best shown in our Fig.\ref{Fig4_bis} which represents the numerical and graphical visualization of the Theorem of Uniqueness. The solution of the system we build up behaves asymptotically, so no mixing-length/time is required. It is required only to wait that amount of time for which, within a given layer, the solution becomes stable to the required precision (in our case the yellow strip of Fig.\ref{Fig4_bis}).

The whole system of ADEs is further simplified to an algebraic system by decoupling the evolution of the generalized coordinates of the
radius and position of a convective element. This result is achieved by means of a series of theorems, corollaries and lemmas that permit the analysis of the different mathematical and physical aspects of the problem, always retaining
the necessary rigour to trace  progress to the final result. The new theory applied to the external convection in the Sun has been proven to yield results (convective fluxes, temperature gradients $\nabla$ and $\nabla_e$, velocity and size of the convective elements) as good as those that are currently obtained with the standard MLT upon having calibrated the mixing-length parameter. The size and path of a convective element will change with the position inside  the convective region, the evolution of the star,  i.e. the particular phase under consideration, and finally the stellar mass itself.

We have two final comments. First  we comment briefly on the reasons why it was necessary to develop the theory in the non-inertial reference frame   ${S_1}$  co-moving with the convective element instead of the more natural frame ${S_0}$ co-moving with the star. The flow past a sphere is indeed a well-studied topic of fluid dynamics  (too large to be reviewed here!) and recently the Lagrange formalism has become particularly suitable  to address this kind of problem: see e.g. \citet[][and references therein]{Tuteja2010} for a recent review and discussion. Unfortunately, this approach does not yield  Eq.(\ref{Eq010}) and Eq.(\ref{Eq018}), in the  Theorem  and companion Corollary  discussed in Section \ref{EqcnvQ} which are required to  derive the  acceleration term  in which the properties of the convection element  are related to the depth inside the star. To compute  the kinetic term of  the energy we would require to evaluate the integral  $T = \frac{1}{2}\rho \int_{}^{} {\left\| {{{\bm{v}}_0}} \right\|{d^3}{\bm{x}}} $ which for the potential flow of Eq.(\ref{Eq004}) turns out to  diverge. This would force us to work at the limit condition $\mathop {\lim }\limits_{\xi  \to \infty } {{\bm{v}}_0} = 0$  for Eq.(\ref{Eq005}) and with a suitable potential energy  ${E_P} = {E_P}\left( {z,R} \right)$ in the two generalized coordinates $z$  and $R$ as defined above in ${S_0}$. The resulting Lagrange equations under the approximation of Eq.(\ref{Eq011}) reduce to a system of decoupled equations instead of Eq.(\ref{Eq018}) which in contrast retains the desired coupling between the generalized coordinates. At this point the only viable solution is instead to write a Lagrangian for the non-inertial system, and this is indeed what was derived in \citet{2009A&A...499..385P} (their section 3.1), which represents our starting point.

Second, we compare the new theory with recent statistical analyses of turbulent convection in stars by  \citet[][and references therein]{2014arXiv1401.5176M}. In brief, adopting the so-called Reynolds-Averaged Navier Stokes (RANS) framework in spherical geometry, developed by the authors over the years \citep[e.g.][]{2014arXiv1401.5176M}
they present results for convection occurring in the stellar interiors and evolutionary phases of typical stars. These works represent an ideal tool to set up \textit{numerical } experiments of stellar convection (from 1D to 3D models). However an \textit{analytical }approach provides an understanding of the process in a way that a numerical one does not. In a forthcoming study we will present an extended survey of the impact on stellar models and a direct integration of the ADE system
composed by Eq.(\ref{Eq056}) with Eq.(\ref{Eq018}) to extend the present formalism and theory to the case of overshooting, where the path travelled by the convective element has specifically to be computed \citep[][in preparation]{Pasetto14}.

\begin{appendix}
\section{Mathematical solution of the non-dimensional expansion rate equation}

\begin{figure}%
\includegraphics[width=\columnwidth]{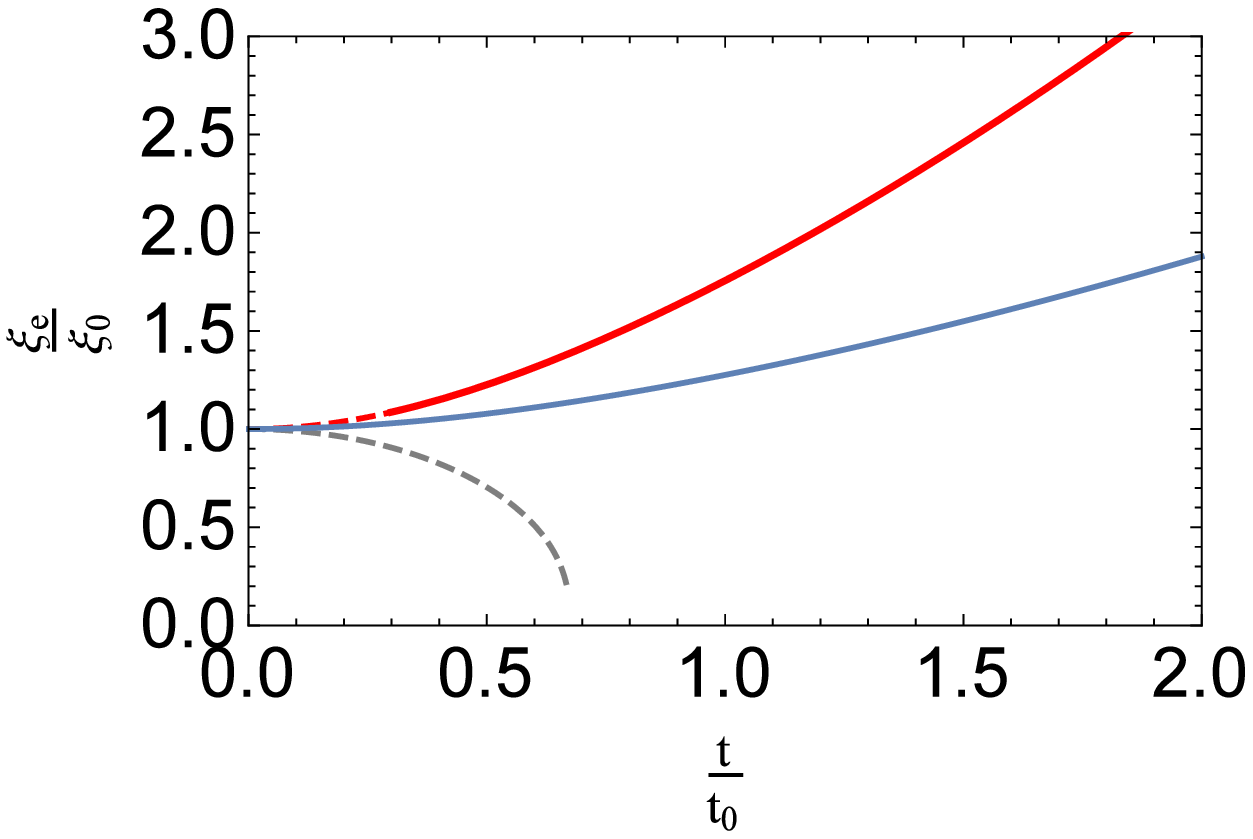}
\caption{Integration of Eq.(\ref{Eq028}), for convection-unstable ($\chi$, thick-red), time averaged ($\bar \chi \left( \tau  \right)$, thick-blue) and stable layers (dashed grey). The red line is partially dashed to remind the reader that Eq.(\ref{Eq028}) holds only on its asymptotic expansion, say for $t > {\hat t}$ for an arbitrary chosen ${\hat t}=0.3 t/t_0$ in the figure (see Section \ref{EqCnv} for detailed discussion) with monotonic character evident from Eq.(\ref{Eq031}). The black semi-circle over $\left\{ {\chi ,\tau } \right\} = \left\{ {1.0,0.0} \right\}$ is excluded.
\label{Fig2}}%
\end{figure}

Despite its elegance, Eq.(\ref{Eq024}) that we report here:
\begin{equation}\label{Eq024appendix}
\chi \frac{{{d^2}\chi }}{{d{\tau ^2}}} + \frac{3}{2}{\left( {\frac{{d\chi }}{{d\tau }}} \right)^2} - 2\chi  = 0,
\end{equation}
is a \textit{non-linear} DE, and so there are no general techniques available in the literature  to solve it.
Nevertheless, as the ODE Eq.(\ref{Eq024}) does not contain explicitly the independent variable, $\tau $, a convenient change of variables is performed by introducing $\frac{{d\chi }}{{d\tau }} = \eta $. In our case we have $\frac{{{d^2}\chi }}{{d{\tau ^2}}} = \frac{{d\eta }}{{d\tau }} = \frac{{d\eta }}{{d\chi }}\frac{{d\chi }}{{d\tau }} = \frac{{d\eta }}{{d\chi }}\eta $. Therefore,  Eq.(\ref{Eq024}) becomes:

\begin{equation}\label{Eq025}
\chi \eta \frac{{d\eta }}{{d\chi }} + \frac{3}{2}{\eta ^2} - 2\chi  = 0,
\end{equation}
which is a lower-order DE, the solution of which is simply:
\begin{equation}\label{Eq026}
\eta \left( \chi  \right) =  \pm \frac{{\sqrt {{c_1} + {\chi ^4}} }}{{{\chi ^{3/2}}}}.
\end{equation}
Using the original variable $\frac{{d\chi }}{{d\tau }} = \eta $ and Eq.(\ref{Eq020}) we obtain
\begin{equation}\label{Eq027}
\frac{{d\chi }}{{d\tau }} =  \pm \frac{{\sqrt {{\chi ^4} - 1} }}{{{\chi ^{3/2}}}},
\end{equation}
where the positive sign is for the expanding convective elements and the negative sign for the contracting ones. The solution of this equation exists for  $\chi  > 1$ strictly.  In the case of an expanding convective element, a solution is always possible. By separating the variables,  we get:
\begin{equation}\label{Eq028}
\int_1^X {\frac{{{\chi ^{3/2}}}}{{\sqrt {{\chi ^4} - 1} }}} d\chi  = \int_0^T {d\tau } ,
\end{equation}
with $X = \frac{{{\xi _e}\left( T \right)}}{{{\xi _0}}}$, whose solution for the LHS  is obtained by means of  two changes of variable, first $y = {\chi^4}$  and then
$z = \frac{{y - 1}}{{{X^4} - 1}}$. We must exclude the case $X=1$ which means that the initial and final sizes are of the convective elements are equal. Therefore in all other cases  $X \ne 1$, the solution is
\begin{equation}\label{Eq029}
\int_1^X {\frac{{{\chi ^{3/2}}}}{{\sqrt {{\chi ^4} - 1} }}} d\chi  = 2{\left( {{X^4} - 1} \right)^{1/2}}{{\kern 1pt} _2}{F_1}\left( {\frac{3}{8},\frac{1}{2};\frac{3}{2};1 - {X^4}} \right),
\end{equation}
where we made use of the standard definition of the Hypergeometrical Function $_2{F_1}\left( {a,b;c,z} \right)$ \citep[e.g.][]{Lebedev}
\begin{equation}\label{Eq030}
_2{F_1}\left( {a,b;c,z} \right) = \frac{{\Gamma \left( c \right)}}{{\Gamma \left( b \right)\Gamma \left( {c - b} \right)}}\int_0^1 {dt\frac{{{t^{b - 1}}{{\left( {1 - t} \right)}^{c - b - 1}}}}{{{{\left( {1 - zt} \right)}^a}}}}.
\end{equation}
The $\Gamma(c)$, $\Gamma(b)$ and $\Gamma(c-b)$ are  the Euler Gamma functions whose values of interest for Eq.(\ref{Eq029}) are $\Gamma \left( 1 \right) = 1$, $\Gamma \left( {{\raise0.5ex\hbox{$\scriptstyle 3$}
\kern-0.1em/\kern-0.15em
\lower0.25ex\hbox{$\scriptstyle 2$}}} \right) = {\raise0.5ex\hbox{$\scriptstyle {\sqrt \pi  }$}
\kern-0.1em/\kern-0.15em
\lower0.25ex\hbox{$\scriptstyle 2$}}$, and $\Gamma \left( {{\raise0.5ex\hbox{$\scriptstyle 1$}
\kern-0.1em/\kern-0.15em
\lower0.25ex\hbox{$\scriptstyle 2$}}} \right) = \sqrt \pi $. This pretty and elegant solution is particularly suitable for numerical implementations thanks to the large body of literature  on the  $_2{F_1}\left( {a,b;c;x} \right)$ functions. Most importantly, the intersection with the solution of the RHS (right-hand side) of Eq.(\ref{Eq028}) can be proven to be \textit{bijective} thus representing an unique solution of the equation for a convecting element in ${S_1}$.
The solution obtained from Eq.(\ref{Eq029}) and Eq.(\ref{Eq028}) for unstable and stable convective regions is plotted in  Fig.\ref{Fig2}.\textit{ We remind the reader that the solution is no longer valid for  small or null values of $\tau $, but only in the limit of very large $\tau$ and hence $t$, mathematically $t \to \infty $}, i.e. only when the condition for Eq.(\ref{Eq011}) is satisfied.  Of course this does not represent a difficulty owing to the large possible freedom that we have on the time scale ${t_0}$ (say, seconds, days, to years over the timescale of the stellar existence, from Myr to Gyr).

\end{appendix}

\label{lastpage}

\section{Acknowledgments}
We thank the referee, Prof. Georges Meynet, for careful report and suggestions. SP thanks D. Crnojevic for her kind hospitality at the Royal Observatory of Edinburgh where part of this work was done. CC and EKG acknowledge support by the Collaborative Research Center ``The Milky Way System'' (SFB 881) of the German Research Foundation (DFG) at Heidelberg University, particularly in the framework of sub-project A5 and in the SFB visitor program.

\begin{small}
\bibliographystyle{mn2e}                    
\bibliography{MNRAS_Biblio}                 
\end{small}
\end{document}